%% file: strobel.tex
\documentclass[12pt]{article}
 
\usepackage{epsfig}

\title{Neutron star properties in the \\ Thomas-Fermi model}

\author{K. Strobel\,$^{a,}$\thanks{e-mail:
    kstrobel@gsm.sue.physik.uni-muenchen.de}\,, F.
  Weber\,$^{a,}$\thanks{Present address (until 31 December 1996): Lawrence
    Berkeley National Laboratory, Nuclear Science Division, MS 70A~-~3307,
    Berkeley, CA 94720, USA}\,, M. K. Weigel\,$^{b}$, \\ and Ch.
  Schaab\,$^{a}$\\ $^{a}$ Institut  f\"ur Theoretische Physik, Universit\"at
  M\"unchen\\ Theresienstrasse 37, D-80333 M\"unchen, Germany \\ $^{b}$ Sektion
  Physik, Universit\"at M\"unchen\\ Am Coulombwall 1, D-85748 Garching, Germany
}
\begin{document}
\baselineskip15pt
\maketitle
\begin{abstract}
  The modern nucleon-nucleon interaction of Myers and Swiatecki, adjusted to
  the properties of finite nuclei, the parameters of the mass formula, and the
  behavior of the optical potential is used to calculate the properties of
  $\beta$--equilibrated neutron star matter, and to study the impact of this
  equation of state on the properties of (rapidly rotating) neutron stars and
  their cooling behavior.  The results are in excellent agreement with the
  outcome of calculations performed for a broad collection of sophisticated
  nonrelativistic as well as relativistic models for the equation of state.
\end{abstract}


LBNL--39604 \\

\newpage
\section{Introduction}

The equation of state (EOS) of dense nuclear matter plays a decisive role in
many branches of physics. It is an important input quantity for the
understanding of heavy--ion collisions, supernova explosions, and the structure
of neutron stars \cite{OPP}-\cite{BRO96}. Therefore great efforts have been
made by several groups to derive the EOS of such matter.  These models can be
divided into two categories, namely nonrelativistic potential models and
relativistic, fieldtheoretical ones.  A frequently used model that belongs to
the first category is the Skyrme Hamiltonian which has the advantage, being a
phenomenological approach, to reproduce the nuclear data rather well. Extensive
calculations of neutron star matter properties for this scheme have been
performed (see, for instance, Refs.\,\cite{PRA,LAM,LAT}). More elaborate,
microscopic schemes constitute the nonrelativistic potential models, where one
uses realistic nucleon-nucleon potentials in combination with 
sophisticated many--body approximations. A typical example of this kind is the
work of Wiringa et al.  \cite{WIR}, in which the properties of matter were
calculated for a Hamiltonian containing two-- and three--body interactions.
These extensive calculations with some later refinements are still the basis
for the calculation of neutron star properties (in the nonrelativistic
approach) in several investigations \cite{BAL}. The disadvantage of such a
microscopic treatment is the numerical complexity of the method. For that
reason it is very tempting to use simpler models which are easier to deal with,
and make comparisons with respect to the properties of finite nuclei, the
parameters of the mass formula, and neutron stars. For this purpose we selected
the new Thomas--Fermi approach (TF) of Myers and Swiatecki \cite{MYE90}, where
the density-- and momentum dependent interaction is given by [the upper (lower)
sign corresponds to nucleons with equal (unequal) isospin]:
\begin{eqnarray}
\label{I.1}
v_{12\tau} &= & - \frac{2T_{0_\tau}}{\rho_0}
\,f\,\left(\frac{r_{12}}{a}\right) \\ \nonumber
 && \times \left(\frac{1}{2}(1 \mp \xi) \alpha - \frac{1}{2}(1 \mp \zeta)
\left(\beta \left(\frac{p_{12}}{p_0}\right)^2 - \gamma \frac{p_0}{|p_{12}|}
+ \sigma \left(\frac{2 \bar \rho}{\rho_0}\right)^{\frac{2}{3}}\right)\right)~.
\end{eqnarray} The quantities $\rho_0$, $p_0$, and $T_{0\tau} (=p_0^2/2m_\tau)$
denote the baryon number density, Fermi momentum, and the kinetic
single--particle energy of symmetric nuclear matter at saturation,
respectively. The potential's radial dependence, $f$, is chosen to be Yukawa
type.  The choice $\xi \not= \zeta$ leads to a better description of asymmetric
nuclear systems, and the behavior of the optical potentials is improved by the
term $\sigma (2\bar\rho/\rho_0)^{2/3}$, where $\overline{\rho}^{2/3} = 0.5
(\rho_{1}^{2/3} + \rho_{2}^{2/3})$.  As discussed in great detail by Myers and
Swiatecki, this model (with only seven free parameters) has the advantage over
the former Seyler--Blanchard interaction \cite{SEY,MYE69} to not only reproduce
the ground--state properties of finite nuclei and symmetric infinite nuclear
matter but also the optical potential and, as revealed by a comparison with the
theoretical investigations of Friedman and Pandharipande \cite{FRI}, the
properties of pure neutron matter, too.  Therefore we think this approximation,
which is able to account for so many features in a very satisfying fashion,
constitutes, despite its simplicity, an excellent candidate for the
investigation of dense matter in the nonrelativistic approach (see also
\cite{Hartmann} where a simpler force has been used).  Since the standard
comparison with nuclear physics quantities has already been performed by Myers
and Swiatecki \cite{MYE90}, we will concentrate in this contribution on the
properties of neutron star matter and the structure and thermal evolution of
neutron stars constructed for their model.  The outcome will be compared with
the one obtained for a broad collection of other models for the equation of
state, which comprises nonrelativistic potential models as well as
relativistic, fieldtheoretical ones.  We shall not go into details concerning
these models, whose properties have been discussed in numerous investigations
prior to this one (see, for instance, Refs.\,\cite{WEB93,PRA,BRO96} and
\cite{SER}-\cite{HUB93}).

\section{Equation of State}

\subsection{Symmetric and Asymmetric Nuclear Matter}

The explicit expressions for the energy per baryon, E/A, chemical potentials,
$\mu$, pressure, $P$, effective masses, $m^*$, and symmetry energy $J$ are
given in Refs.\,\cite{MYE90,STR} and will not be repeated here again.  For the
parameters of the TF model of Myers and Swiatecki we use the most recent set
denoted TF \cite{MYE..} (We denote the different parameter sets with TF90,
TF94, and TF96. If not stated otherwise the latest parametrization, TF96
($\equiv$ TF), is used.). Earlier parametrizations (TF90, TF94) have larger
statistical errors in fitting the nuclear data. The EOSs for the parameter sets
of Ref.~\cite{MYE..} are rather similar, but the EOS for the parameter set TF90
of Ref.~\cite{MYE90} is stiffer than the latter EOSs (see Table~I).  For the
purpose of comparison, we give the parameters of symmetric infinite nuclear
matter for the different models in Table~I.  (An overview of all EOSs used in
this work is given in Table~II).

Nonrelativistic Brueckner--Hartree--Fock calculations (BHF) with two--body
forces are in general not capable to reproduce the correct saturation
properties. The resulting EOSs are too soft. This deficiency is corrected by
adding three--body forces (TBF) which lead to improved saturation properties
and a stiffer EOS.  Nevertheless the BHF calculations occur to be not
satisfactory enough for the description of symmetric infinite nuclear matter
(see Refs.\,\cite{WIR,BAL} and Table~I) and we will therefore not discuss them
in greater detail.  Stiffer EOSs are also obtained in the
Dirac--Brueckner--Hartree--Fock approach (DBHF)
\cite{HUB93},\cite{BRO90}-\cite{HUB96a}. The general trend seems to be that
variational calculations, BHF supplemented with TBF, and DBHF calculations
agree with each other more or less only up to moderate densities.  At higher
densities one obtains a closer agreement of DBHF with the variational
calculations, depending on the matter's asymmetry \cite{WIR,BAL,MAC,SUM}. Our
main intention is to discuss the properties of the EOS of neutron star matter,
and compare the results with those obtained in the nonrelativistic many-body
approach as well as the fieldtheoretical treatment.  As examples belonging to
the first category, we have selected EOSs from the investigations of Wiringa et
al.  \cite{WIR}, which have been used frequently in neutron star calculations
by others, too. The two samples are (cf.\ Table~II) the variationally-based
EOSs computed for the Urbana UV14 two--nucleon potential supplemented with the
UVII three--nucleon potential (WUU), and the density--dependent UV14 potential
supplemented with the three--nucleon interaction TNI (WUT).  As representatives
of the DBHF and relativistic Hartree--Fock (RHF) scheme we have chosen EOSs
from Refs.\,\cite{WEB89,HUB93,BRO90,SUM}.

As pointed out above, the DBHF--EOSs are stiffer than BHF--EOSs based on
two--body forces, but by introducing three--body forces one can simulate the
relativistic behavior to a certain extent \cite{WIR,BAL}.  In Fig.~1 we
illustrate the situation for symmetric nuclear matter.  The agreement of all
models, with the exception of the BHF model based on two--body forces
\cite{WIR,BAL,SUM,HUB96a}, is better than in the case of pure neutron matter
(see Figs.~2 and 3), where larger deviations occur above $\rho =
0.4~\mbox{fm}^{-3}$\,.  The potential model of Myers and Swiatecki leads to a
pure neutron--matter EOS which is located between the two nonrelativistic
variational calculations WUU and WUT.  The larger differences between
nonrelativistic and relativistic models for higher asymmetry are explainable by
the different behavior of the symmetry energy. The nonrelativistic potential
models cause a stronger repulsion in isospin singlet states for higher
densities \cite{WIR}, which makes for increasing asymmetry the nonrelativistic
EOSs softer in comparison with the relativistic ones. This behavior can be
clearly seen in the density dependence of the symmetry energy (see Fig.~4),
where TF shows a similar behavior as WUT. It seems to be a general feature
that relativistic treatments lead to a monotonic increase of the symmetry
energy, in contrast to the nonrelativistic models where the symmetry energy
saturates (for BHF \cite{WIR,BAL,SUM}), or even decreases with increasing
density (variational calculations).  As we shall discuss later, this feature
has a severe impact on the composition of neutron star matter, since the
symmetry energy determines the proton fraction of neutron star matter. The
proton fraction in turn plays an important role for the cooling behavior of
neutron stars, since proton fractions above $\sim$ 0.11 -- 0.13 permit stars to
cool very efficiently via the direct Urca process \cite{Boguta81a,Lattimer91}.

\subsection{Neutron Star Matter}

Since neutron stars are bound by gravity, which is much weaker than the Coulomb
force, and have life times practically infinite 
compared to the characteristic weak interaction time scale,
$\tau_w \sim 10^{-10}~{\rm s}$, neutron star matter is subject to the
constraints of charge neutrality and generalized $\beta$--equilibrium,
respectively \cite{RUD,WEB93,PRA,WIR,GLE,STR}:
\begin{equation}
\label{1}
\sum_{B = p,n} q_B~ (2 J_B +1)~ \frac{k_{F,B}^3}{\pi^2} -
    \sum_{L = e,\mu} \frac{k^3_{F,L}}{3\pi^2}  = 0 ~, 
\end{equation}
and
\begin{equation}
\label{2}
\mu_B = \mu_n - q_B\; \mu_L ~, \qquad \mu_\mu = \mu_e ~ .
\end{equation} Here $\mu_B(q_B)$ denotes the chemical potentials (electric
charges) of baryons, and $\mu_L$ the chemical potentials of leptons $e^-$,
$\mu^-$.  Due to these constraints the calculation of neutron--star--matter
properties differs from the treatment of asymmetric matter, since the
composition has to be determined selfconsistently subject to the two additional
constraints (\ref{1}) and (\ref{2}).  The new baryon/lepton degrees of freedom
lower the energy and pressure of neutron star matter in comparison with pure
neutron matter \cite{WEB93,PRA,GLE,WEB89,HUB96a,HUB96}. This feature is
illustrated in Fig.~5, where the pressure of neutron star matter is shown as
function of density and compared with the pressure of pure neutron matter. Due
to the new degrees of freedom, the pressure is lower in $\beta$--stable matter.
To demonstrate the influence of muons on the EOS, we also show the EOS computed
for $n,~p,$ and $e^-$ only. It indicates that the muons play only a minor role
for the stiffness of the EOS. Nevertheless they are important for the
composition.  For the neutron star matter EOS computed for the model of Myers
and Swiatecki one does expect deviations from the microscopic calculations
which are of the same order as for the calculations performed for fixed
asymmetry.

In Figs.~6 and 7 we show the energy per baryon of neutron star matter for the
different models.  The nonrelativistic models give in general softer EOSs, but
the model of Myers and Swiatecki comes rather close to the relativistic results
(HWW1), especially at lower densities.  The earlier parametrization of Myers
and Swiatecki, that is TF\,90, comes even closer to the relativistic outcome
\cite{STR}, since the EOS is much stiffer.

The same behavior holds also for the pressure, which is exhibited in Fig.~8.
Much larger differences than for the EOS occur for the composition of neutron
stars \cite{STR}.  Due to the behavior of the symmetry energy discussed above,
one encounters in relativistic models a monotonous increase of the proton
fraction, $x$ $(\equiv \rho_p/\rho)$ with density (in the case of no hyperons
and/or meson condensates), in contrast to nonrelativistic microscopic theories
and the TF model of Myers and Swiatecki for which $x$ decreases (or saturates
as for BHF) at higher densities.  This is illustrated in Fig.~9 and Table~III,
where the composition is compared for several different models. In this context
we note that the stiffness of modern fieldtheoretical models which -- besides
neutrons and protons account for the population of more massive baryon states
(especially hyperons) too -- comes closer to the stiffness one of
nonrelativistic neutron star matter EOSs \cite{WEB93,GLE,WEB89,HUB96}.  Here
one obtains again a decrease of the electron/muon fraction, since it is
energetically favorable for the system to achieve charge neutrality among the
hyperons themselves.

\section{Neutron Star Properties}

In order to determine neutron star properties one has to solve Einstein's
equations, for which knowledge of the EOS, i.e.\ $P(\epsilon)$, is necessary.
Einstein's curvature tensor $G_{\mu\nu}$ ($R_{\mu\nu},~ g_{\mu\nu},~ R$, and
$T_{\mu\nu}$ denote the Ricci tensor, metric tensor, Ricci scalar, and
energy--momentum tensor density, respectively) is determined by 
\begin{equation}
\label{III.1}
G_{\mu\nu} \equiv R_{\mu\nu} - \frac{1}{2}\; g_{\mu\nu} \; R = 8
\pi\,T_{\mu\nu}\left(\epsilon,P(\epsilon)\right)~.
\end{equation} For the determination of the properties of (rapidly) rotating
neutron stars and their limiting periods, one has to generalize the
Schwarzschild metric of a spherically symmetric, static star to the one of a
rotationally deformed, axially symmetric body \cite{FRI86,WEB92}:
\begin{eqnarray}
\label{III.2}
ds^2 & = & e^{-2\nu(r,\theta;\Omega)} dt^2 + e^{2\psi(r,\theta;\Omega)} 
(d\phi - \omega(r,\theta;\Omega) dt)^2 \\ \nonumber
& & +~e^{2\mu(r,\theta;\Omega)} d\theta^2 
+~e^{2\lambda(r,\theta;\Omega)} dr^2~.
\end{eqnarray} $\Omega$ denotes the neutron star's rotational frequency, and
$\omega(r,\theta;\Omega)$ is the angular velocity of the local inertial frames
(dragging of the local inertial frames) \cite{WEB93,FRI86,WEB91}).

Of great interest with respect to the identification of rapidly spinning
pulsars as rotating neutron stars is the maximum possible rotational frequency
of such an object, for which no simple stability criteria exist in general
relativity. However an absolute upper limit is given by the Kepler frequency,
$\Omega_K$, at which mass sedding at the equator sets in.  It is determined as
the solution of the following transcendental equation \cite{HUB94,FRI86,WEB92},
\begin{equation}
\label{III.3}
\Omega_K = \left\{e^{\nu(r,\theta;\Omega_K) - \psi(r,\theta;\Omega_K)}
  V(r,\theta,\Omega_K) + \omega(r,\theta,\Omega_K)\right\}_{\rm{eq}}~,
\end{equation} which is to be evaluated at the star's equator. $V$ denotes the
velocity of a particle rotating at the star's equator.  Equations (\ref{III.1})
and (\ref{III.3}) are to be solved simultaneously by a selfconsistent numerical
iteration scheme, since neither the metric functions in (\ref{III.2}) nor the
frequencies $\Omega_K$ and $\omega$ (all depend on the star's unknown mass) are
known.

In Fig.~10 we exhibit the neutron star masses as a function of central energy
density for different EOSs. Since WUU and WUT behave rather similar up to mass
densities of about $\sim 10^{15}$g/cm$^3$, the mass-density curves computed for
these two EOSs are very similar to each another.  The deviations beyond that
density arise from the different three--body forces. The TF--EOS behaves for
smaller energy densities like the relativistic EOS, but shows at higher
densities features similar to a medium--stiff EOS.  So the properties of
neutron stars computed for TF and HWW2, which are listed in Tables~IV and V,
are similar to each other as long as the star's mass is below about
$1.4~M_\odot$. Besides HWW2 the EOS of Brockmann and Machleidt \cite{BRO90,SUM}
has been chosen here as a further representative of a relativistic EOS.  The
former accounts for the presence of hyperons in neutron star matter too
\cite{HUB94,HUB96}.  The stiffness of an EOS at high densities is known to play
a key role for the maximum possible mass that a neutron star can have. Since
the nonrelativistic EOSs tend to be less stiff at high densities than their
relativistic counterparts (the only exception is WUU), we obtain less massive
neutron stars for the nonrelativistic EOSs.

Another effect caused by the stiffness of an EOS concerns the radius of the
stellar configurations. In general, the stiffer the EOS the lower the star's
central mass density and thus the lower the gravitational force that pulls the
star together.  So stars constructed for stiffer EOSs will have larger radii
than those constructed for softer EOSs. This behavior is confirmed in Fig.~11,
where the radius--mass relations of neutron stars constructed for TF and HWW2
are shown: the stiffer model for the EOS, that is HWW2, leads to somewhat
bigger stars, provided their mass is beyond about one solar mass. The masses of
stars lighter than this value are determined by the stiffness/softness of the
EOS at nuclear and intermediate densities where TF behaves stiffer than HWW2,
leading to neutron stars that are somewhat bigger for TF than for HWW2.  The
radial changes due to rotation at $\Omega_K$ amount about 2 km, except for the
very light stars of each sequence.  The mass increase of the heaviest stars due
to rapid rotation is about $0.3~M_\odot$ (see also Fig.~12).  Various
properties of static as well as rotating neutron stars, computed for a number
of different EOSs, are listed in Tables~VI--VIII. All properties refer to
neutron stars of mass $M=1.4~M_\odot$ (static and rotating), about which
the observed masses tend to scatter.

In Fig.~13 we show the density profiles of nonrotating neutron stars of mass
$M=1.4~M_\odot$, which reflect what has just been discussed above in connection
with the dependence of stellar radii on the stiffness/softness of the EOS: The
microscopic EOSs WUU and WUT, behaving somewhat softer than TF and HWW2, lead
to smaller stars with a higher central density (for more phenomenological
relativistic EOSs, see Ref.~\cite{SCHWEB}). Moreover, as can be inferred from
Fig.~13, the TF--EOS density profile is rather close to the one of the
relativistic HWW2--EOS because of the rather similar behavior of these EOSs
over the energy densities relevant for a star of this mass.

A final remark concerns the velocity of sound, which does not exeed the causal
limit in the TF model for the range of relevant neutron star densities, i.e.,
$c_{s} \leq 0.96~c$ for $\rho \leq 1.246~\rm{fm}^{-3}$. The causal limit is
reached at $\rho = 1.36~\rm{fm}^{-3}$, which is considerably higher than the
highest density reached in the most massive star of the sequence.

As mentioned above, the Kepler frequency gives an absolute upper limit on the
critical rotational frequency of a neutron star. Its rotation may also be
limited by the gravitational radiation--reaction instability. Since the theory
of this effect is too lengthy, we will not outline the treatment in this
investigation but refer to \cite{WEB93,WEB91,WEB93IOP} for details. In Fig.~14
we exhibit the limiting rotational periods set by both the Kepler criterion and
the gravitational radiation--reaction instability. The latter depend on
temperature, for which we have chosen representative values of $T=10^6$~K,
corresponding to an old neutron star, and $10^{10}$~K typical for a young newly
formed one.  The limiting periods set by the gravitation--radiation instability
are larger than the Kepler periods and thus set a more stringent limit on rapid
rotation than mass shedding. Moreover they increase with temperature, since
viscosity damps the instability modes less efficiently in hot stars.  A
comparison between TF and HWW2 shows that the differences in rotational periods
are rather small, except for hot neutron stars where somewhat larger deviations
show up.  The shaded rectangle in Fig.~14 covers masses and periods of observed
pulsars \cite{NAG}. Hence both EOSs, based on the assumption that neutron star
matter is made up of hadrons and leptons rather than other kinds of exotic
forms of matter \cite{WEB93}, are in accordance with the body of presently
existing data.

A further important facet of neutron star physics concerns the cooling behavior
of such objects. In Ref.~\cite{SCHWEB} we have already calculated and discussed
in great detail their cooling behavior for different EOSs depending on the
involved processes (see, for instance, Table~5 of Ref.~\cite{SCHWEB}).  Here we
show and compare the cooling tracks of different neutron star models of mass
$M=1.4~M_\odot$.  The underlying EOSs are the nonrelativistic TF and WUU
models, and the relativistic HWW2. As an example of an enhanced neutrino
emission process we consider the direct Urca process, which is only possible if
the proton fraction exceeds a certain critical value ($\approx 0.13$, see Refs.
\cite{Boguta81a,Lattimer91}). Among the considered EOSs, only HWW2 allows for
the direct Urca process. This leads to a large drop of the star's surface
temperature at $\tau\approx30$~yr (see the dashed curve of the nonsuperfluid
model in Fig.~15), since the core cools down very fast via the enhanced
neutrino emission process in comparison with the crust, causing a temperature
inversion in young stars. Depending on the crust thickness, the cooling wave
formed by the temperature gradient reaches the surface and causes the sharp
decrease of the surface temperature, followed by a rather flat behavior of the
cooling tracks up to $10^7$ years \cite{SCHWEB,SCHVOS}. We show also the
behavior of superfluid neutron stars. Superfluidity reduces the neutrino
emissivity, the heat capacity, and the thermal conductivity by an exponential
factor of $\exp(-\Delta/kT)$, where $\Delta$ denotes the gap energy (see
Table~4 of Ref.~\cite{SCHWEB} for the used gap energies).

The observational data are described in Ref. \cite{SCHHER} (see Table~IX).  The
obtained effective surface temperature depends crucially on whether a hydrogen
atmosphere is used or not. Since the photon flux, measured solely in the X-ray
energy band, does not allow to determine what atmosphere one should use, we
consider both the blackbody model (solid error bars in Fig.~15) and the
hydrogen-atmosphere model (dashed error bars). The plotted errors represent the
$3\sigma$ error range due to the small photon fluxes. The pulsars ages are
determined by their spin-down times assuming a canonical value of 3 for the
breaking index. In reality the breaking index may be quite different from 3.
Its variation between 2 and 4, for instance, would change the age of Geminga as
indicated by the horizontal error bar shown at the bottom of Fig.~15.

The cooling tracks of the nonrelativistic models, that is TF and WUU, are
almost identical. Since the relativistic HWW2 model cools mainly through the
direct Urca process in the neutrino cooling era, its cooling behavior differs
considerably from the nonrelativistic models. Superfluidity reduces the
neutrino emissivity which leads to a higher surface temperature in the neutrino
cooling era. Later, in the photon cooling era, the surface temperature becomes
smaller for the superpfluid models.  The observational data can almost
perfectly be described by both the nonsuperpfluid and the superfluid
nonrelativistic models as well as the superfluid relativistic model, provided
one assumes that the pulsars have no hydrogen atmosphere (except PSR 1055-52,
which could be explained by internal heating; see, e.g., Refs.
\cite{VanRiper95a,Schaab96a}).  However, if some of the pulsars prove to have a
hydrogen atmosphere, these models seem to be too hot. In this case superfluid
enhanced cooling or intermediate cooling \cite{SCHVOS} might be necessary.
Whether the observed pulsars have a hydrogen atmosphere could be decided if one
considers multiwavelength observations, as suggested by Pavlov et al.
\cite{Pavlov96a}.

\section{DISCUSSION AND CONCLUSIONS}

We have calculated within the modern nonrelativistic TF model of Myers and
Swiatecki the properties of neutron star matter and compared the outcome with
microscopic nonrelativistic and relativistic treatments. The main purpose of
this contribution was to test this model with respect to its implications for
the physics of neutron stars.  It turned out that, despite its simplicity, the
modern TF model shows the same features with respect to the EOS of neutron star
matter as complicated nonrelativistic variational calculations. Moreover the
model is in even closer, over the relevant density regions encountered in
neutron stars, with microscopic relativistic EOSs. The composition of neutron
star matter obtained for TF is similar to other nonrelativistic models for
which the symmetry energy saturates or even decreases at higher densities too,
favoring so, in contrast to relativistic models, again an increase of the
neutron fraction at higher densities. The resulting gross structural properties
of neutron stars (masses, redshifts etc.)  are in accordance with observations.
Also interesting is the finding that the limiting rotational neutron star
periods computed for TF, set by either the Kepler criterion or the
gravitational--radiation reaction instability, agree rather closely with the
results of modern relativistic EOSs. The limiting period of a $M\sim
1.4~M_\odot$ neutron star is about 1~ms, which too is in agreement with the
present observations.  Limiting periods of this size seem to be a general
feature of EOSs computed hadronic/leptonic matter \cite{WEB93,HUB94}. Finally
we studied the thermal evolution of neutron stars for TF. Here, somehow
surprisingly, it turned out that no enhanced cooling mechanism needs to be
invoked to achieve agreement between theory and observation; cooling via only
the modified Urca process occurs to be sufficient.  (We recall that cooling via
the direct Urca process is not permitted for TF, as is the case for the
nonrelativistic microscopic neutron star matter EOSs stdied in this work.)

From these findings we conclude that the modern Thomas--Fermi model of Myers
and Swiatecki is not only very suitable for the description of low--energy
nuclear physics but for phenomena in high--density regime too! Our study
reveals that the TF--EOS of neutron star matter shows similar features as
sophisticated microscopic nonrelativistic EOSs and, over some density range,
comes even close to microscopic relativistic EOSs. The neutron star properties
and the cooling behavior derived for TF are in excellent agreement with the
body of presently existing data.  Due to these surprising features we think
that the model of Myers and Swiatecki is also a very good candidate for the
nonrelativistic (many--body) treatment of neutron stars and, therefore, should
be included in further nonrelativistic investigations in this field.  The
handling of this model with only seven parameters is relatively simple in
comparison with elaborated microscopic schemes, which makes it an ideal
candidate for this kind of astrophysical studies.

\bigskip
\section*{Acknowledgments:} We would like to thank W.~D.~Myers and 
W.~J.~Swiatecki for clarifying remarks and helpful discussions.

\clearpage
\section*{Table captions}

\begin{description}
\item[TABLE~I:] Saturation density, $\rho_{00}$, energy per baryon, $E/A$,
  incompressibility, $K$, and symmetry energy, $J$, of symmetric infinite
  nuclear matter for different models. These are TF: Thomas--Fermi model of
  Myers and Swiatecki \cite{MYE90,MYE..}; WUU and WUT: variational calculation
  performed for UV14 plus UVII and UV14 plus TNI, respectively \cite{WIR,BAL};
  BHF2 and BHF3: Brueckner--Bethe approximation with two-- (AV\,14) and
  three--body forces (AV\,14+UVII), respectively \cite{WIR,BAL}; DBHF$^{(1)}$
  and DBHF$^{(2)}$: relativistic Dirac--Brueckner--Hartree-Fock approximation
  with solutions in the space of positive energy--spinors only \cite{BRO90} and
  the full Dirac space \cite{HUB93}, respectively (A and B denote different
  Brockmann--Machleidt potentials \cite{HUB93,BRO90}); NL1 and NL--SH:
  phenomenological relativistic mean field theory for the parameter sets of
  Reinhard and Sharma,
  respectively \cite{HUB93}); SKM$^*$ and SIII: Skyrme forces \cite{BRA}.

\item[TABLE~II:] Equations of state (EOSs) used in this work.

\item[TABLE~III:] Proton fraction, $x$, and energy per baryon (in MeV) of
  beta--stable matter (neutrons, protons, electrons and muons).

\item[TABLE~IV:] Neutron star properties for beta--stable TF model
  (neutrons, protons, electrons and muons). $M_G (\epsilon_c)$ denotes the
  gravitational mass in solar mass units as a function of central mass density.
  The central star pressure is denoted $p_c$.  The amu mass $M_A$ minus the
  gravitational mass $M_G$ is effectively the binding energy liberated when the
  NS is formed. $R$ stands for the star's radius, $\Delta_c$ denotes the
  stellar crust using 2.4$\times 10^{14}$ g\,cm$^{-3}$ as the boundary, $I$
  denotes the moment of inertia, and $z$ the surface redshift.  For a
  comparison with WUU and WUT see Ref.~\cite{WIR}.

\item[TABLE~V:] Same as Table IV, but for the HWW2 model.

\item[TABLE~VI:] Comparison of the properties of a static, spherical NS of mass
  1.4 M$_{\odot}$ for different models.

\item[TABLE~VII:] Properties of rotating neutron star models of mass
  $M=1.4~M_{\odot}$ and rotational period $P=1$~ms, calculated for different
  EOSs.  The entries are (from top to bottom): central energy density,
  $\epsilon_c$; percental mass increase relative to nonrotating star model of
  same $\epsilon_c$, $\Delta M/M$; equatorial and polar radii, $R_{\rm eq}$ and
  $R_{\rm p}$, respectively; moment of inertia, $I$; stability parameter, $t$;
  injection energy, $\beta$; redshift at the pole, $z_{\rm p}$; eccentricity,
  $e$; quadrupole moment, $\Pi$. The influence of different parametrizations of
  the TF model on NS properties is shown too.  The stiffness of the TF EOSs
  decreases from TF90n (pure neutron matter, represents the stiffest EOS) to
  TF96.

\item[TABLE~VIII:] Same as Table~VII, but for a rotational period of $P=1.6$~ms.

\item[TABLE~IX:] Surface temperatures, $T_s^\infty$, and spin--down ages, $t$,
  of several observed pulsars \cite{SCHWEB,SCHVOS,SCHHER}.
\end{description}

\clearpage

\begin{table}
\begin{center}
TABLE I \\[1cm]
\begin{tabular}
{|l|l|l|l|l|}
\hline
Method & E/A & $\rho_{00}$ & $K$ & J \\
       & (MeV) & (fm$^{-3}$)   & (MeV) & (MeV) \\
\hline
TF90 & -16.53 & 0.165 & 301 & 32 \\
TF94 & -16.04 & 0.161 & 234 & 32 \\
TF96 & -16.24 & 0.161 & 234 & 33 \\
WUU & -15.5 & 0.175 & 202 & 30 \\
WUT\,$^{\rm a)}$ & -16.6 & 0.157 & 261 & 29 \\
BHF\,2 & -17.8 & 0.280 & 247 & -- \\
BHF\,3 & -15.2 & 0.194 & 209 & -- \\
DBHF\,A$^{(1)}$ & -15.6 & 0.185 & 290 & -- \\
DBHF\,A$^{(2)}$ & -16.5 & 0.174 & 280 & 34 \\
DBHF\,B$^{(1)}$ & -13.6 & 0.174 & 249 & -- \\
DBHF\,B$^{(2)}$ & -15.7 & 0.172 & 249 & 33 \\
NL\,1 & -16.4 & 0.152 & 212 & 43 \\
NL--SH & -16.3 & 0.146 & 356 & 36 \\
SkM$^*$ & -15.8 & 0.160 & 216 & 30 \\
SIII & -15.9 & 0.145 & 355 & 28 \\
\hline
a) in Ref.\,\cite{SUM}: & -15.7 & 0.158 & 246 &\\
\hline
\end{tabular}
\end{center}
\end{table}

   \begin{table}
\begin{center}
TABLE II \\[1cm]
\begin{tabular}
{|l|l|l|l|}
\hline
EOS & Interaction  & Many-body & Ref.\\
    &              & approach  &     \\
\hline
TF90 & Myers and Swiatecki & Thomas-Fermi &  \cite{MYE90,STR} \\
TF94 & Myers and Swiatecki & Thomas-Fermi &  \cite{MYE..} \\
TF96 & Myers and Swiatecki & Thomas-Fermi &  \cite{MYE..} \\
WUU  & Urbana 2-nucleon potential UV14 plus   & variational & \cite{WIR}  \\
     & Urbana 3-nucleon potential UVII        &  &            \\
WUT  & Urbana 2-nucleon potential UV14        & variational & \cite{WIR}  \\
     & Urbana 3-nucleon potential TNI         &  &            \\
BHF2 & Argonne 2-nucleon potential AV14       & nonrelativistic & 
\cite{WIR,BAL}  \\
     &                                        & Brueckner-Bethe &            \\
BHF3 & Argonne 2-nucleon potential AV14 plus  & nonrelativistic & 
\cite{WIR,BAL}  \\
     & Urbana 3-nucleon potential UVII        & Brueckner-Bethe &            \\
RB   & Brockmann-Machleidt potential A & relativistic & \cite{BRO90,SUM} \\
     &                                 & Brueckner- &    \\
     &                                 & Hartree-Fock  &  \\
HWW1 & Brockmann-Machleidt potential A & relativistic & \cite{HUB96a}  \\
     & (no hyperons)                   & Brueckner- &    \\
     &                                 & Hartree-Fock  &  \\
HWW2 & Brockmann-Machleidt potential B & relativistic & \cite{HUB94,HUB96}  \\
     & + HFV                           & Brueckner- &    \\
     &                                 & Hartree-Fock  &  \\
HV  & Exchange of $\sigma$, $\omega$ and & relativistic & \cite{WEB93}  \\
     & $\rho$ mesons (with hyperons)     & Hartree &    \\
HFV  & Exchange of $\sigma$, $\omega$ and & relativistic & \cite{WEB93}  \\
     & $\rho$ mesons (with hyperons)      & Hartree-Fock &    \\
\hline
\end{tabular}
\end{center}
\end{table}
   

   \begin{table}
   \begin{center}
   TABLE III\\[1cm]
   \begin{tabular}{|l|r|r|r|r|r|r|}
   \hline
       & \multicolumn{2}{c|}{TF}&\multicolumn{2}{c|}{WUU}&\multicolumn{2}{c|}
       {WUT}\\
   \cline{2-7}
   \raisebox{0.0ex}[0cm][0cm]{$\rho$ (fm$^{-3}$)}
     &  $x(\rho)$ & $E(\rho,x)$ & $x(\rho)$ & $E(\rho,x)$ & $x(\rho)$ 
 & $E(\rho,x)$  \\
   \hline
   0.07 &  0.0276 & 5.88 & 0.0193 & 8.13 & 0.0247 & 6.25 \\
   \hline
   0.08 & 0.0308 & 6.25 & 0.0213 & 8.66 & 0.0273 & 6.43 \\
   \hline
   0.10 &  0.0365 & 7.05 & 0.0253 & 9.70 & 0.0317 & 6.87 \\
   \hline
   0.125 &  0.0427 & 8.25 & 0.0300 & 11.06 & 0.0357 & 7.69 \\
   \hline
   0.15 &  0.0502 & 9.48 & 0.0345 & 12.59 & 0.0388 & 8.89 \\
   \hline
   0.175 &  0.0572 & 10.97 & 0.0402 & 14.18 & 0.0418 & 10.32  \\
   \hline
   0.20 &  0.0630 & 12.83 & 0.0464 & 15.92 & 0.0442 & 12.13  \\
   \hline
   0.25 &  0.0715 & 17.78 & 0.0572 & 20.25 & 0.0469 & 16.76  \\
   \hline
   0.30 &  0.0767 & 24.36 & 0.0632 & 25.78 & 0.0476 & 22.53 \\
   \hline
   0.35 &  0.0794 & 32.57 & 0.0673 & 32.60 & 0.0476 & 29.18 \\
   \hline
   0.40 &  0.0803 & 42.34 & 0.0741 & 40.72 & 0.0472 & 36.75 \\
   \hline
   0.50 &  0.0780 & 66.28 & 0.0854 & 61.95 & 0.0401 & 56.06 \\
   \hline
   0.60 &  0.0725 & 95.56 & 0.0959 & 90.20 & 0.0311 & 79.19 \\
   \hline
   0.70 &  0.0651 & 129.62 & 0.1108 & 126.20 & 0.0195 & 106.04 \\
   \hline
   0.80 &  0.0572 & 167.97 & 0.1215 & 170.50 & 0.0054 & 135.46 \\
   \hline
   1.00 & 0.0417 & 256.15 & 0.1239 & 291.10 & 4.8e-4 & 200.89 \\
   \hline
   1.25 &  0.0274 & 385.57 & 0.0855 & 501.00 & 0.0000 & 294.00 \\
   \hline
   1.50 &  0.0152 & 534.64 & 0.0195 & 753.00 & 0.0000 & 393.00 \\
   \hline
   \end{tabular}
   \end{center}
   \end{table}

   
   \begin{table}
   \begin{center}
   TABLE IV\\[1cm]
   \begin{tabular}{|c|c|c|c|c|c|c|c|}
   \hline
   $\epsilon_{c}$ & $p_{c}$        & $M_{G}$       & $M_{A} - M_{G}$ & $R$  
 & $\Delta_{c}$ & $I$         & $z$ \\
   (10$^{14}$)       & (10$^{34}$)    & ($M_{\odot}$) & ($M_{\odot}$)   
 & (km) & (km)         & (10$^{44}$) &     \\
   (g/cm$^{3}$)      & (dyn/cm$^{2}$) &               &                 & 
     &              & (g~cm$^{2}$) &     \\
   \hline
    2.5 &     0.2703 & 0.131 & 0.001 & 17.89 & 16.08 &  0.569 & 0.011 \\
   \hline
    3.0 &     0.4496 & 0.184 & 0.003 & 14.17 &  9.90 &  0.854 & 0.020 \\ 
   \hline
    3.5 &     0.7009 & 0.248 & 0.005 & 12.68 &  7.06 &  1.26 & 0.030 \\
   \hline
    4.0 &     1.046 & 0.327 & 0.009 & 11.99 &  5.36 &  1.82 & 0.043 \\
   \hline
    5.0 &     2.007 & 0.504 & 0.022 & 11.53 &  3.54 &  3.30 & 0.072 \\
   \hline
    6.0 &     3.387 & 0.696 & 0.041 & 11.45 &  2.57 &  5.15 & 0.104 \\
   \hline
    7.0 &     5.218 & 0.887 & 0.067 & 11.45 &  1.98 &  7.21 & 0.139 \\
   \hline
    8.0 &     7.518 & 1.07 & 0.098 & 11.45 &  1.59 &  9.26 & 0.174 \\
   \hline
   10.0 &    13.46 & 1.36 & 0.163 & 11.38 &  1.13 & 12.9 & 0.244 \\
   \hline
   12.5 &    23.28 & 1.62 & 0.239 & 11.19 &  0.82 & 16.0 & 0.322 \\
   \hline
   15.0 &    35.38 & 1.78 & 0.297 & 10.95 &  0.64 & 17.9 & 0.387 \\
   \hline
   17.5 &    49.47 & 1.88 & 0.339 & 10.71 &  0.53 & 18.7 & 0.442 \\
   \hline
   20.0 &    65.27 & 1.94 & 0.366 & 10.47 &  0.46 & 18.9 & 0.487 \\
   \hline
   25.0 &   101.1 & 1.99 & 0.392 & 10.06 &  0.37 & 18.6 & 0.553 \\
   \hline
   30.0 &   141.6 & 2.00 & 0.396 &  9.73 &  0.33 & 17.9 & 0.598 \\
   \hline
   35.0 &   185.6 & 1.99 & 0.390 &  9.43 &  0.27 & 17.0 & 0.633 \\
   \hline
   \end{tabular}
   \end{center}
   \end{table}
   
   
   \begin{table}
   \begin{center}
   TABLE V\\[1cm]
   \begin{tabular}{|c|c|c|c|c|c|c|c|}
   \hline
   $\epsilon_{c}$ & $p_{c}$        & $M_{G}$       & $M_{A} - M_{G}$ & $R$  
& $\Delta_{c}$ & $I$         & $z$ \\
   (10$^{14}$)       & (10$^{34}$)    & ($M_{\odot}$) & ($M_{\odot}$)   
& (km) & (km)         & (10$^{44}$) &     \\
   (g/cm$^{3}$)      & (dyn/cm$^{2}$) &               &                 &  
    &              & (g~cm$^{2}$) &     \\
   \hline
    2.5 &     0.3146 & 0.144 & 0.002 & 14.47 & 12.44 &  0.598 & 0.015 \\
   \hline
    3.0 &     0.5156 & 0.212 & 0.004 & 12.59 &  8.03 &  1.04 & 0.026 \\ 
   \hline
    3.5 &     0.7731 & 0.282 & 0.007 & 11.95 &  6.11 &  1.54 & 0.037 \\
   \hline
    4.0 &     1.055 & 0.345 & 0.010 & 11.68 &  5.06 &  2.02 & 0.047 \\
   \hline
    5.0 &     2.218 & 0.539 & 0.024 & 11.37 &  3.29 &  3.66 & 0.078 \\
   \hline
    6.0 &     3.655 & 0.735 & 0.048 & 11.36 &  2.41 &  5.59 & 0.112 \\
   \hline
    7.0 &     5.724 & 0.955 & 0.085 & 11.44 &  1.83 &  8.09 & 0.152 \\
   \hline
    8.0 &     9.006 & 1.21 & 0.146 & 11.52 &  1.37 &  11.4 & 0.204 \\
   \hline
   10.0 &    17.48 & 1.61 & 0.280 & 11.55 &  0.91 & 17.0 & 0.302 \\
   \hline
   12.5 &    30.81 & 1.91 & 0.423 & 11.42 &  0.64 & 21.5 & 0.404 \\
   \hline
   15.0 &    46.21 & 2.06 & 0.519 & 11.20 &  0.50 & 23.6 & 0.481 \\
   \hline
   17.5 &    61.29 & 2.13 & 0.571 & 11.00 &  0.43 & 24.2 & 0.530 \\
   \hline
   20.0 &    78.08 & 2.17 & 0.602 & 10.81 &  0.40 & 24.2 & 0.568 \\
   \hline
   25.0 &   112.3 & 2.19 & 0.624 & 10.45 &  0.33 & 23.4 & 0.619 \\
   \hline
   30.0 &   148.3 & 2.18 & 0.621 &  10.18 &  0.31 & 22.2 & 0.647 \\
   \hline
   35.0 &   185.4 & 2.15 & 0.609 &  9.95 &  0.29 & 21.1 & 0.665 \\
   \hline
   \end{tabular}
   \end{center}
   \end{table}


    \begin{table}
    \begin{center}
    TABLE VI\\[1cm]
    \begin{tabular}{|l|c|c|c|c|}
    \hline
    Quantity  & TF & WUU & WUT & HWW2 \\
    \hline
    $\epsilon_{c} (10^{14}$ g/cm$^{3})$ & 10.30 & 10.42 & 12.12 & 8.9\\
    \hline
    $P_{c} (10^{34}$ dyn/cm$^{2})$         & 14.54 & 15.74 & 18.58 & 12.37 \\
    \hline
    $M_{G}/M_\odot$                        & 1.4   & 1.4   & 1.4  & 1.4\\
    \hline
    $(M_{A} - M_{G})/M_\odot$              & 0.173 & 0.162 & 0.169 & 0.203 \\
    \hline
    $ R$ (km)                              & 11.37 & 11.15 & 10.86 & 11.56\\
    \hline
    $\Delta_{c}$ (km)                      & 1.08  & 1.06  & 0.88 & 1.12 \\
    \hline
    $I (10^{44}$ g~cm$^{2}$)               & 13.34 & 12.80 & 12.30  & 13.97\\
    \hline
    $z$                                    & 0.254 & 0.261 & 0.271  & 0.248\\
    \hline
    \end{tabular}
    \end{center}
    \end{table}


\begin{table}
\begin{center}
TABLE VII\\[1cm]
\begin{tabular}{|l|rrrrrr|} \hline
Quantity               &TF90n     & TF90    &TF94     &TF96     &HWW2      
&WUU \\
\hline
$\epsilon_c~({{\rm MeV/fm}^3})$   
                       &410.6     &430.1    &490.3    &508.1    &451.3     
&531.2    \\
$\Delta M/M$           &17        &13       &13       &13       &14        &11 
      \\
$R_{\rm eq}$~(km)      &14.50     &13.59    &12.99    &12.82    &13.03     
&12.02    \\
$R_{\rm p}$~(km)       &10.64     &10.35    &10.21    &10.16    &10.18     
&9.82    \\
${\rm log}\;I/({{\rm g~cm}^2})$ 
                       &45.0829  &45.0704  &45.0568  &45.0525  &45.0596   
&45.0345  \\
$t$                    &0.082     &0.076    &0.068    &0.066    &0.070     
&0.060   \\
$\beta$                &0.610     &0.600    &0.595    &0.593    &0.594     
&0.579    \\
$z_{\rm p}$            &0.2792    &0.2905   &0.2966   &0.2986   &0.2979    
&0.3141   \\
$e$                    &0.68      &0.65     &0.62     &0.61     &0.62      
&0.58     \\
$\Pi$~(km$^3$)         &14.7      &12.6     &10.4     &9.8      &11.0 &8.1  \\
\hline
\end{tabular}
\end{center}
\end{table}

\begin{table}
\begin{center}
TABLE VIII\\[1cm]
\begin{tabular}{|l|rrrrrr|} \hline
Quantity               &TF90n     &TF90     &TF94     &TF96     &HWW2     &WUU 
\\
\hline
$\epsilon_c~({{\rm MeV/fm}^3})$   
                       &454.0     &465.8    &531.3    &549.9    &480.9     
&560.7    \\
$\Delta M/M$           &6         &5        &5        &5        &6         &4  
     \\
$R_{\rm eq}$~(km)      &13.05     &12.47    &12.00    &11.88    &12.07     
&11.30   \\
$R_{\rm p}$~(km)       &11.74     &11.35    &11.03    &10.94    &11.06     
&10.51     \\
${\rm log}\;I/({{\rm g~cm}^2})$ 
                      &45.1419  &45.1275  &45.1055  &45.0990  &45.1155   
&45.0803  \\
$t$                    &0.031     &0.029    &0.026    &0.025    &0.027     
&0.023   \\
$\beta$                &0.650     &0.635    &0.625    &0.622    &0.625     
&0.606  \\
$z_{\rm p}$            &0.2425    &0.2549   &0.2649   &0.2679   &0.2649    
&0.2841   \\
$e$                    &0.44      &0.42     &0.39     &0.39     &0.40      
&0.37    \\
$\Pi$~(km$^3$)         &5.5       &4.8      &3.9      &3.7      &4.2  &3.1   \\
\hline
\end{tabular}
\end{center}
\end{table}

\begin{table}
\begin{center}
TABLE IX\\[1cm]
\begin{tabular}
{|l|l|l|l|}
\hline
Pulsar & $t$ [yr] & Model atmosphere & $T_s^\infty [K]$ \\
\hline
0833-45 & $1.1 \times 10^4$ & blackbody & $1.3 \times 10^6$ \\
(Vela) & & magnetic H--atmosphere & $7.0^{+1.6}_{-1.3} \times 10^5$ \\
\hline
0656+14 & $1.1 \times 10^5$ & blackbody & $7.8^{+1.5}_{-4.2} \times 10^5$ \\
&  &  H--atmosphere & $5.3^{+1.2}_{-0.9} \times 10^5$ \\
\hline
0630+18 & $3.2 \times 10^5$ & blackbody & $5.2 \pm 3.0 \times 10^5$ \\
(Geminga) &  &  H--atmosphere & $1.7 \pm 1.0 \times 10^5$ \\
\hline
1055--52  & $5.4 \times 10^5$ & blackbody & $7.9^{+1.8}_{-3.0} \times 10^5$
\\
\hline
\end{tabular}
\end{center}
\end{table}

\clearpage
\section*{Figure captions}

\begin{description}
\item[Fig.~1.] EOSs of symmetric infinite nuclear matter. Compared are:
TF, WUU and WUT, BHF2, BHF3, RB, and HWW1. (The abbreviations are
explained in Table II.)

\item[Fig.~2.] Comparison of pure neutron matter EOSs.

\item[Fig.~3.] Enlargement of the EOSs of neutron matter at low densities.
  Labels as in Fig.~1. The triangles (HWW1) show the results of
  DBHF--calculations (potential $A$) performed in the full Dirac space (for
  details, see Ref.~\cite{HUB93,HUB94,HUB96a}).

\item[Fig.~4.] Comparison of the symmetry energy in the relativistic and
  nonrelativistic approaches.

\item[Fig.~5.] Variation of the EOS with composition in the TF--model. Compared
  is pressure versus density of neutron star matter, composed of $p,n,e^-$ and
  $\mu^-$, with the pure neutron matter case.  The influence of muons on the
  EOS is exhibit too.

\item[Fig.~6.] Comparison of EOSs of $\beta$--equilibrated (i.e., $n,p,e^-$ and
  $\mu^-$) neutron star matter.

\item[Fig.~7.] High--density behavior of the EOSs of $\beta$-stable neutron
  star matter.

\item[Fig.~8.] Pressure-density relation of neutron star matter for different
  models.

\item[Fig.~9.] Comparison of the proton fractions associated with the various
  different EOSs studied in this paper.

\item[Fig.~10.] Gravitational mass of static (nonrotating) neutron star
  sequences versus central energy density for different neutron star matter
  EOSs.

\item[Fig.~11.] Neutron star radius versus mass for TF and the
  relativistic HWW2--EOS.  Shown are sequences of stars rotating at their
  general relativistic Kepler frequencies, and at zero frequency.

\item[Fig.~12.] Gravitational neutron star mass versus central star energy
  density for TF and the relativistic HWW2--EOS. Shown are
  sequences of stars rotating at their Kepler frequencies, and at zero
  frequency.

\item[Fig.~13.] Mass density profiles of a $1.4~M_\odot$ neutron star computed
  for several different EOSs.

\item[Fig.~14.] Limiting rotational periods ($P = 2\pi/\Omega$) of neutron star
  sequences versus star mass. Shown are the Kepler periods and the limiting
  periods set by the emission of gravity waves from young neutron stars
  ($10^{10}K$) as well as old ones ($10^{6}K$). The underlying EOSs are
  TF and the relativistic HWW2. The shaded 
  area covers the range of observed periods and masses (see text).
\item[Fig.~15.] Cooling of neutron stars constructed for TF, WUU and HWW2, with
  and without inclusion of superfluidity. The surface temperatures obtained
  with a blackbody-- (magnetic hydrogen) atmosphere are marked by solid
  (short--dashed) error bars (see Table~IX). The uncertainty in the pulsar's
  age is indicated by the error bar in the lower right corner \cite{SCHHER}.
\end{description}

\clearpage

\input{fig.tex}

\end{document}

%% file: fig.tex
   \begin{figure}[tbp] 
   \begin{center}  
   \leavevmode 
   FIGURE 1\\[0.5cm]
   \epsfig{figure=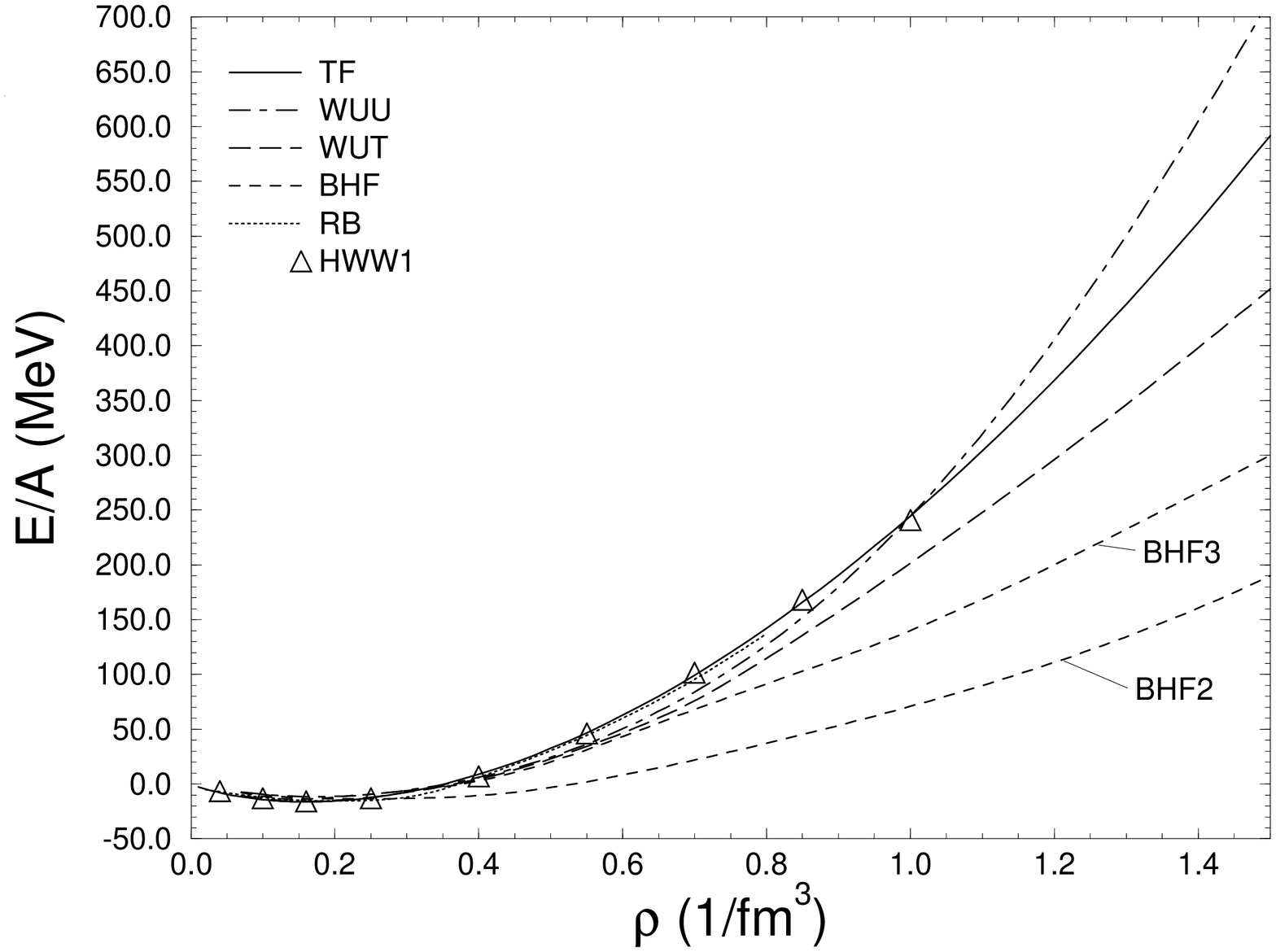,width=70mm,clip=,angle=-90,
           bbllx=95,bblly=35,bburx=560,bbury=680}
   \end{center}
   \end{figure}
   
   \begin{figure}[tbp] 
   \begin{center}  
   \leavevmode 
   FIGURE 2\\[0.5cm]
   \epsfig{figure=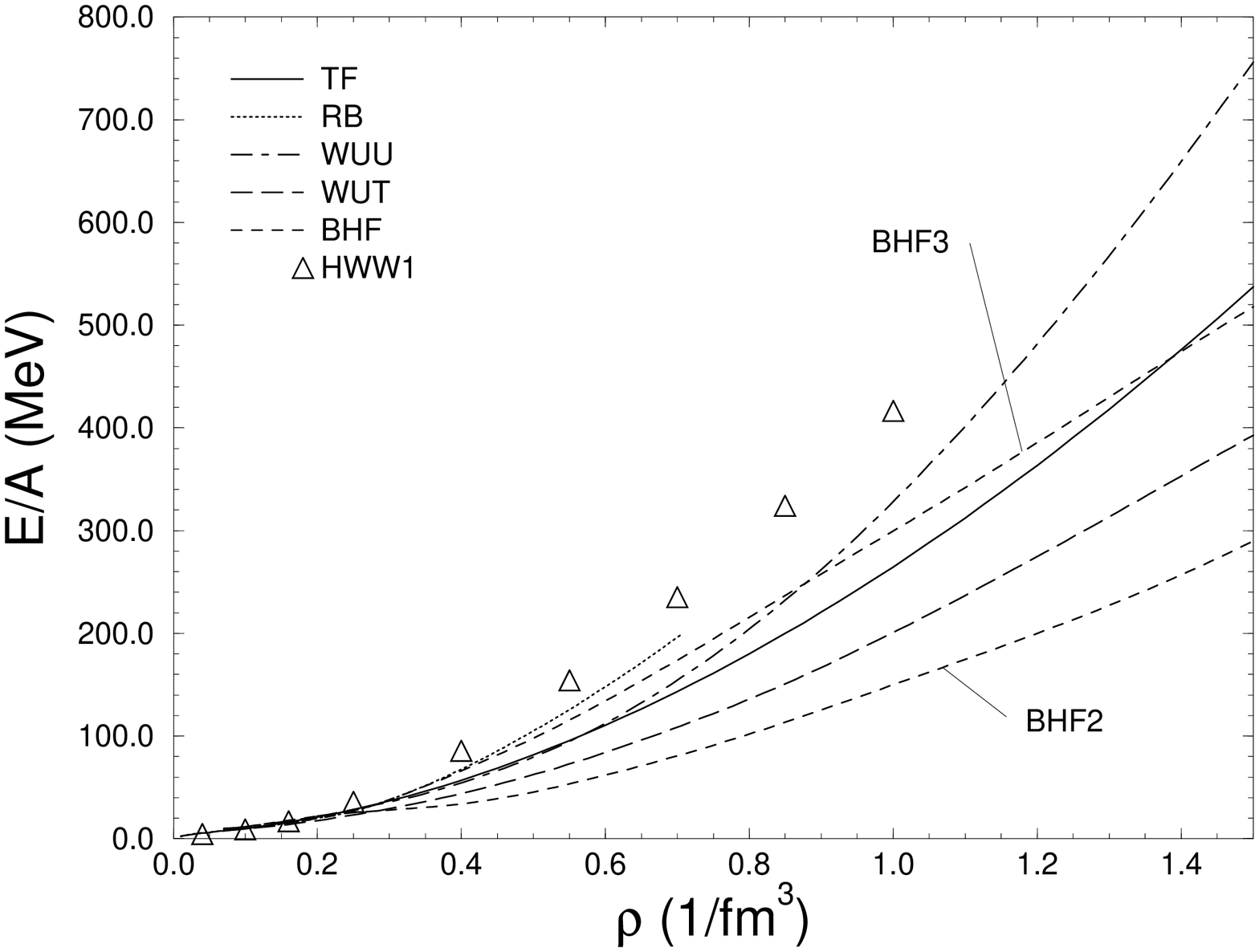,width=70mm,clip=,angle=-90,
            bbllx=95,bblly=35,bburx=560,bbury=680}
   \end{center}
   \end{figure}
   
   \begin{figure}[tbp] 
   \begin{center}  
   \leavevmode
   FIGURE 3\\[0.5cm] 
   \psfig{figure=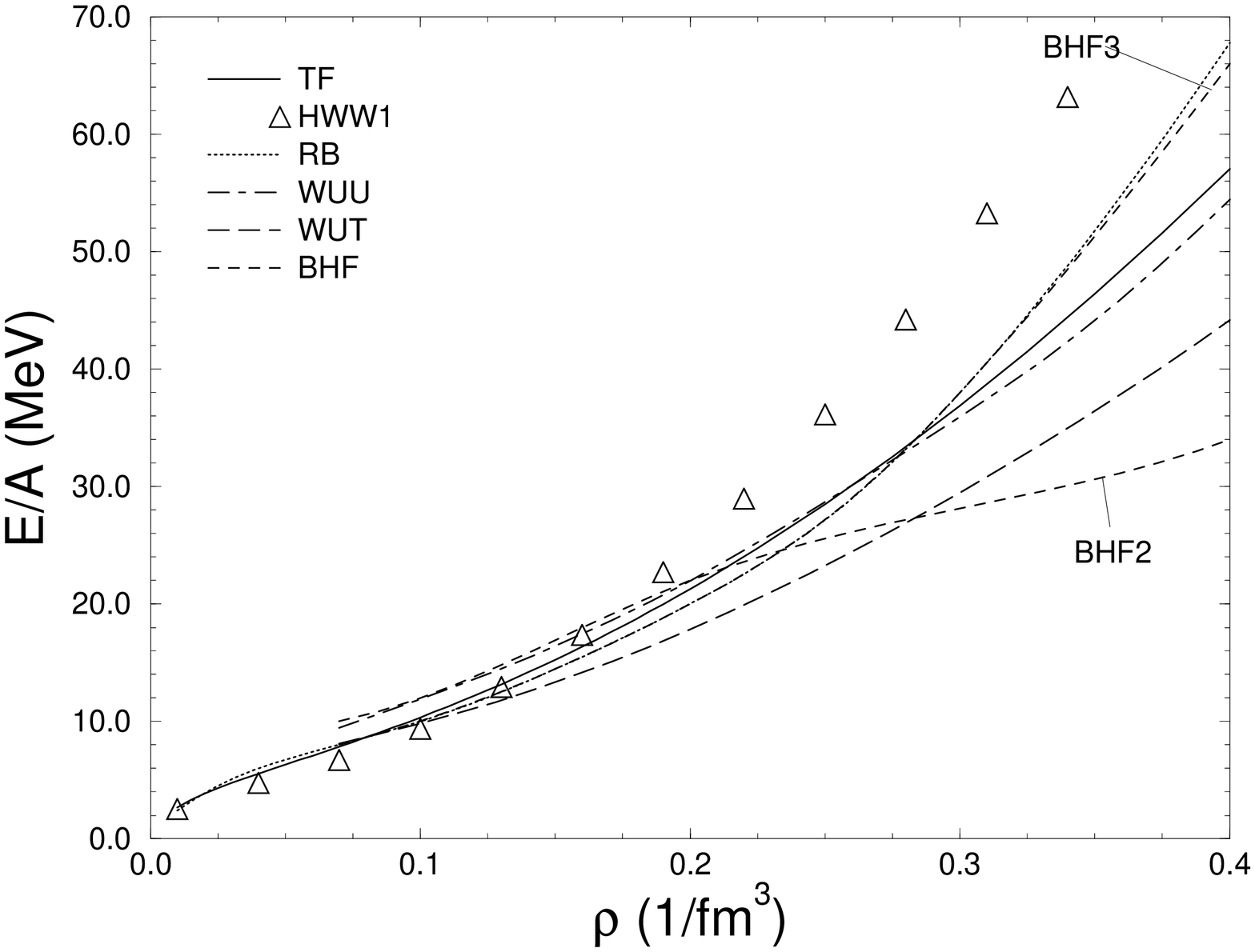,width=70mm,clip=,angle=-90,
          bbllx=95,bblly=35,bburx=560,bbury=680}
   \end{center}
   \end{figure}
   
   \begin{figure}[tbp] 
   \begin{center}  
   \leavevmode 
   FIGURE 4\\[0.5cm]
   \psfig{figure=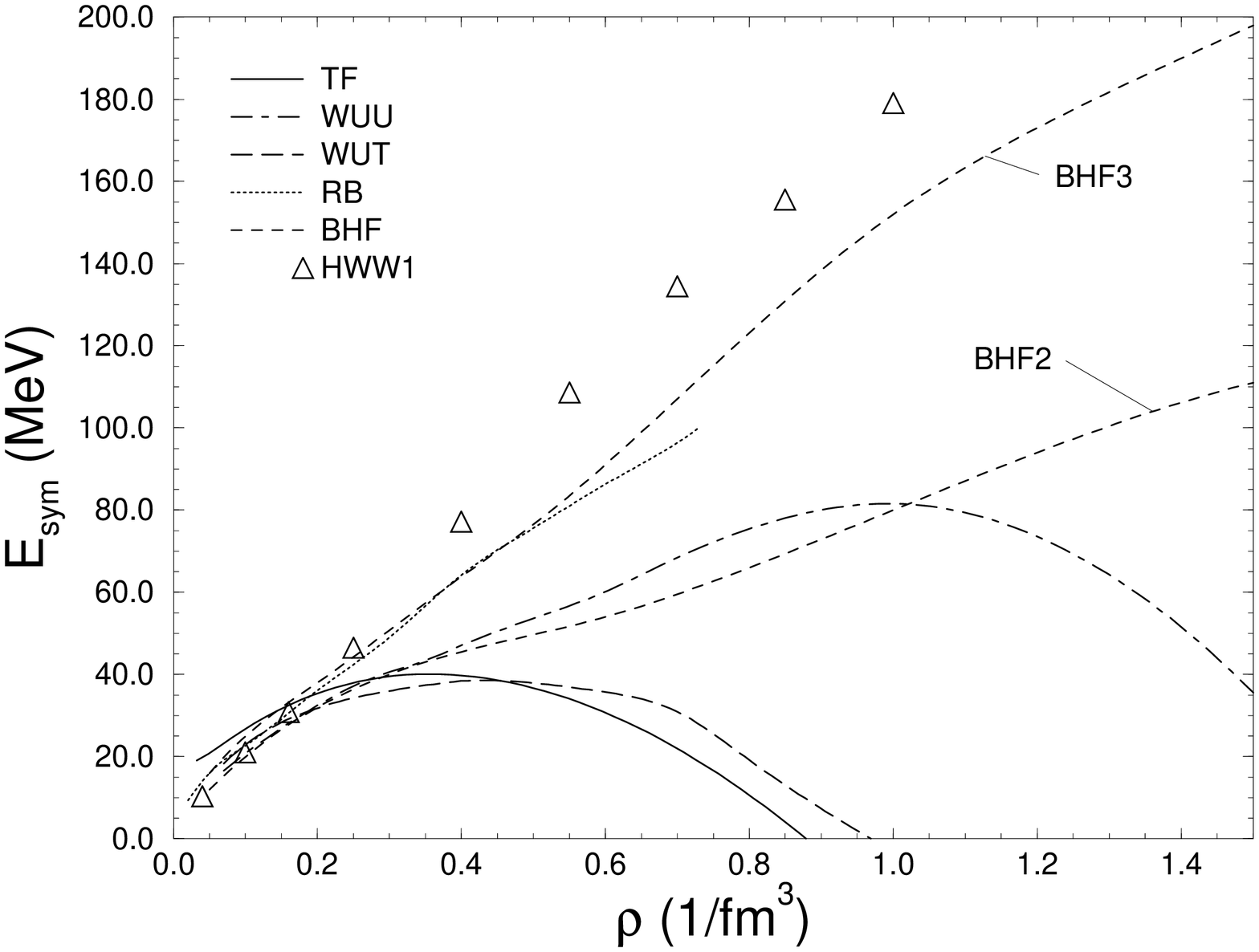,width=70mm,clip=,angle=-90,
          bbllx=95,bblly=35,bburx=560,bbury=680}
   \end{center}
   \end{figure}
   
   \begin{figure}[tbp] 
   \begin{center}  
   \leavevmode 
   FIGURE 5\\[0.5cm]
   \psfig{figure=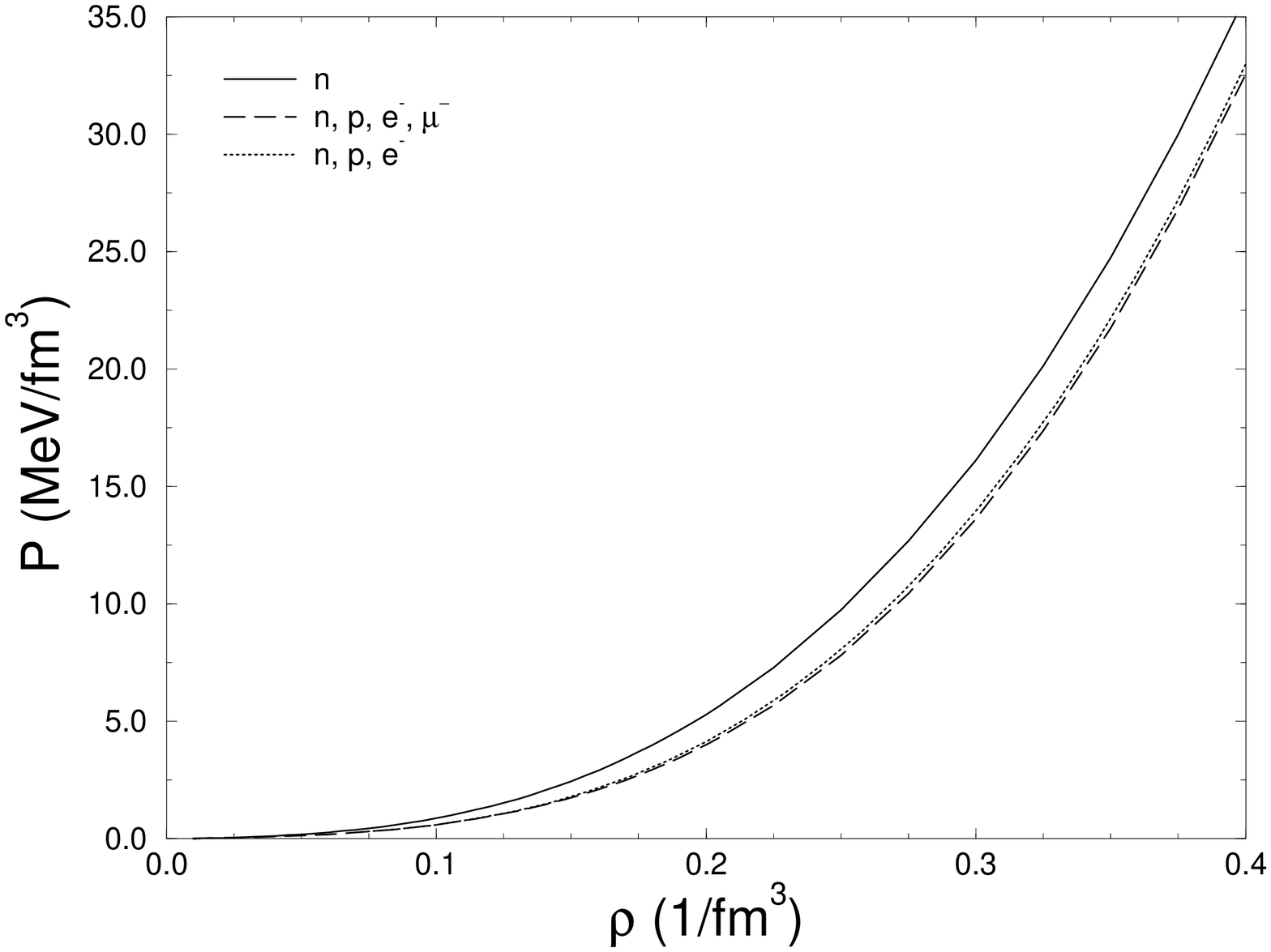,width=70mm,clip=,angle=-90,
           bbllx=95,bblly=35,bburx=560,bbury=680}
   \end{center}
   \end{figure}
   
   \begin{figure}[tbp] 
   \begin{center}  
   \leavevmode 
   FIGURE 6\\[0.5cm]
   \psfig{figure=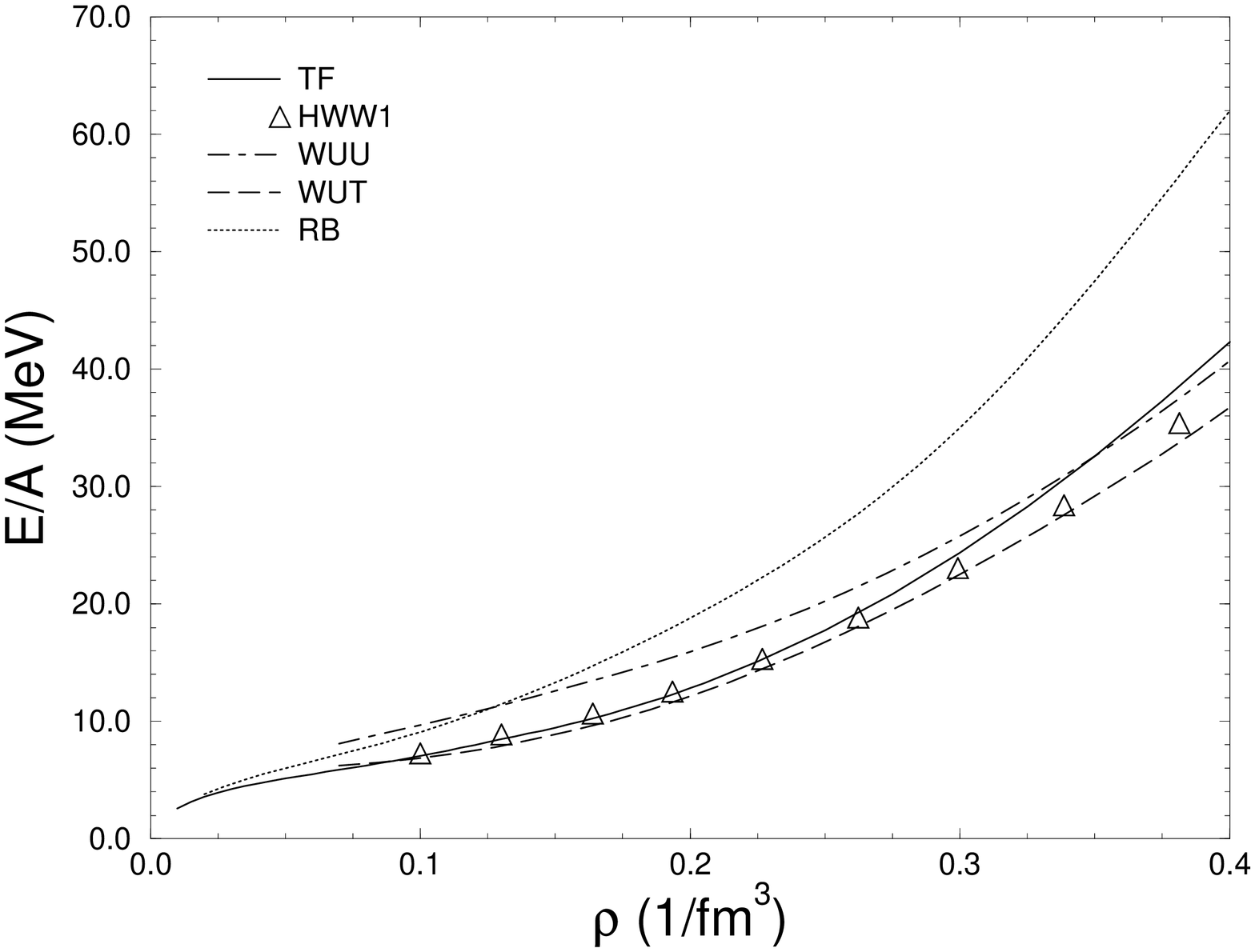,width=70mm,clip=,angle=-90,
           bbllx=95,bblly=35,bburx=560,bbury=680}
   \end{center}
   \end{figure}
   
   \begin{figure}[tbp] 
   \begin{center}  
   \leavevmode 
   FIGURE 7\\[0.5cm]
   \psfig{figure=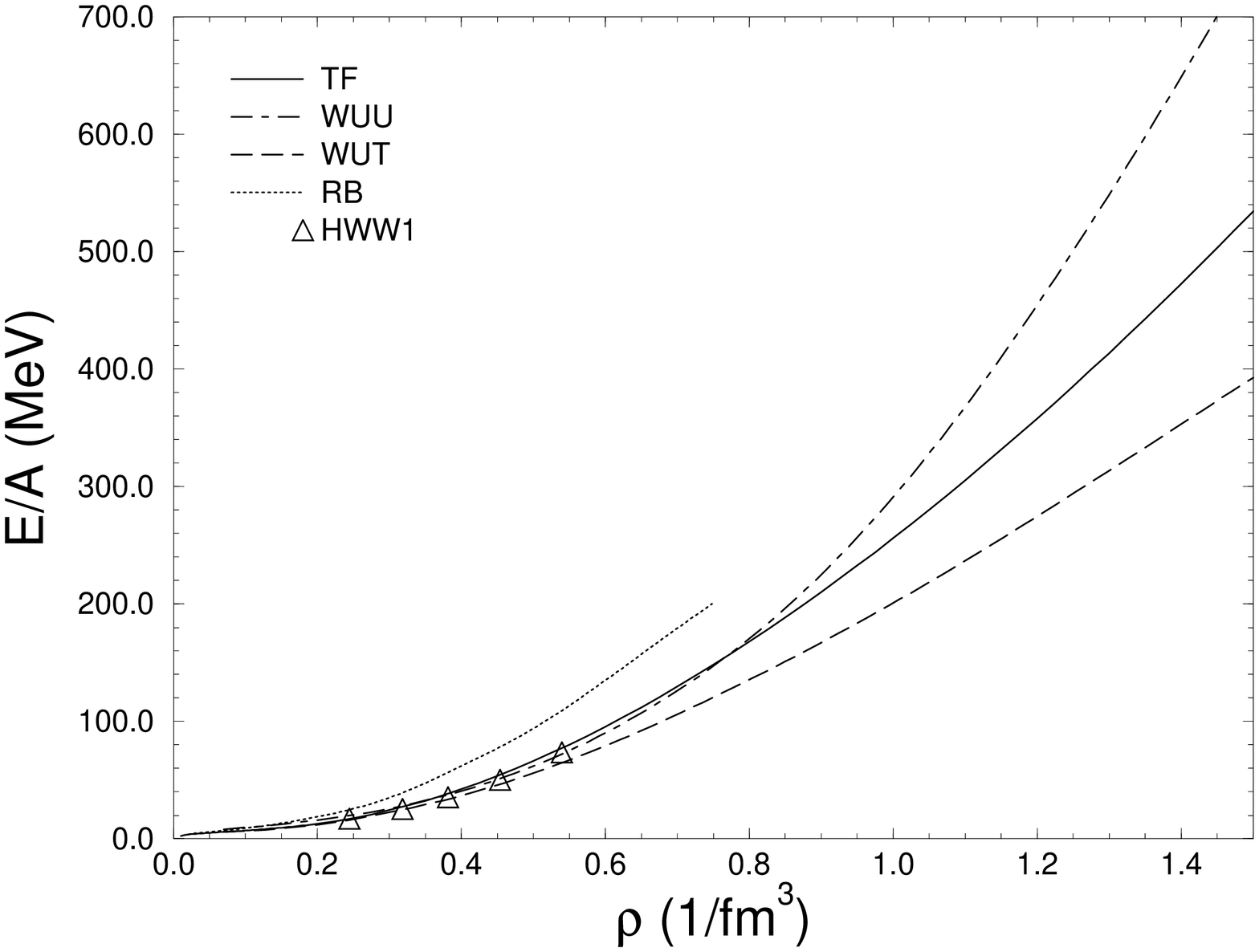,width=70mm,clip=,angle=-90,
           bbllx=95,bblly=35,bburx=560,bbury=680}
   \end{center}
   \end{figure}
   
   \begin{figure}[tbp] 
   \begin{center}  
   \leavevmode 
   FIGURE 8\\[0.5cm]
   \psfig{figure=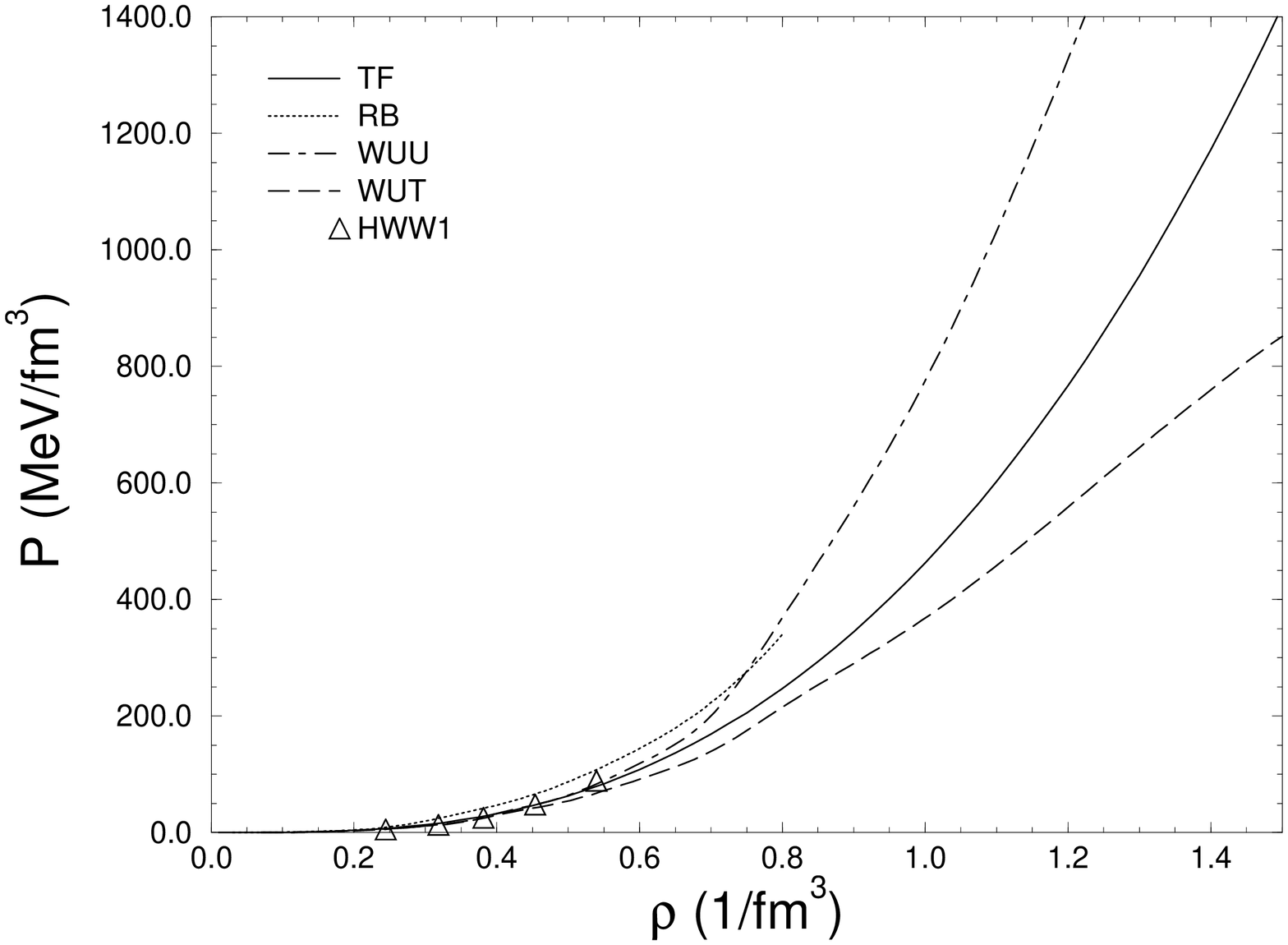,width=70mm,clip=,angle=-90,
           bbllx=95,bblly=25,bburx=560,bbury=680}
   \end{center}
   \end{figure}
   
   \begin{figure}[tbp] 
   \begin{center}  
   \leavevmode 
   FIGURE 9\\[0.5cm]
   \psfig{figure=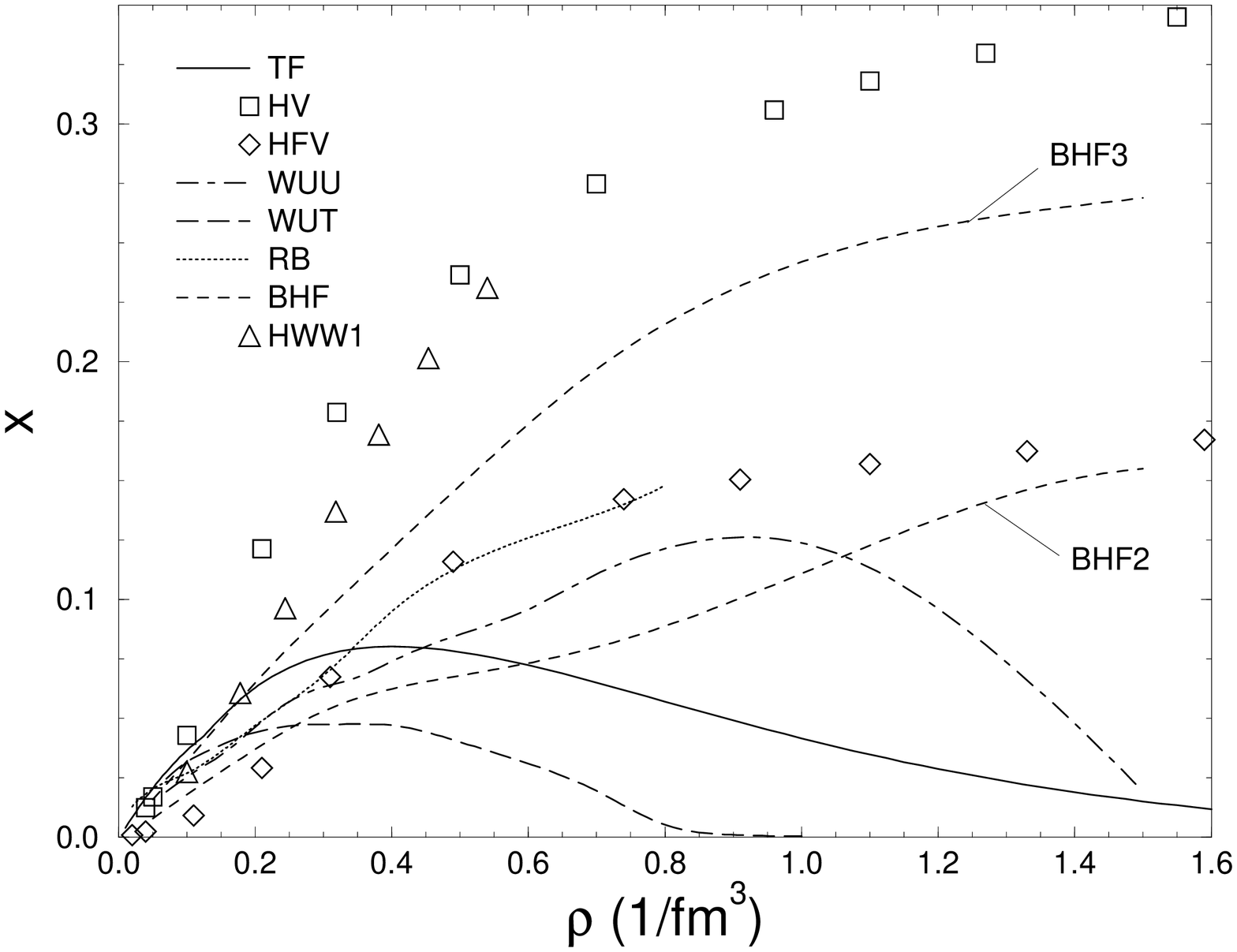,width=70mm,clip=,angle=-90,
          bbllx=95,bblly=35,bburx=560,bbury=680}
   \end{center}
   \end{figure}
   
   \begin{figure}[tbp] 
   \begin{center}  
   \leavevmode 
   FIGURE 10\\[0.5cm]
   \psfig{figure=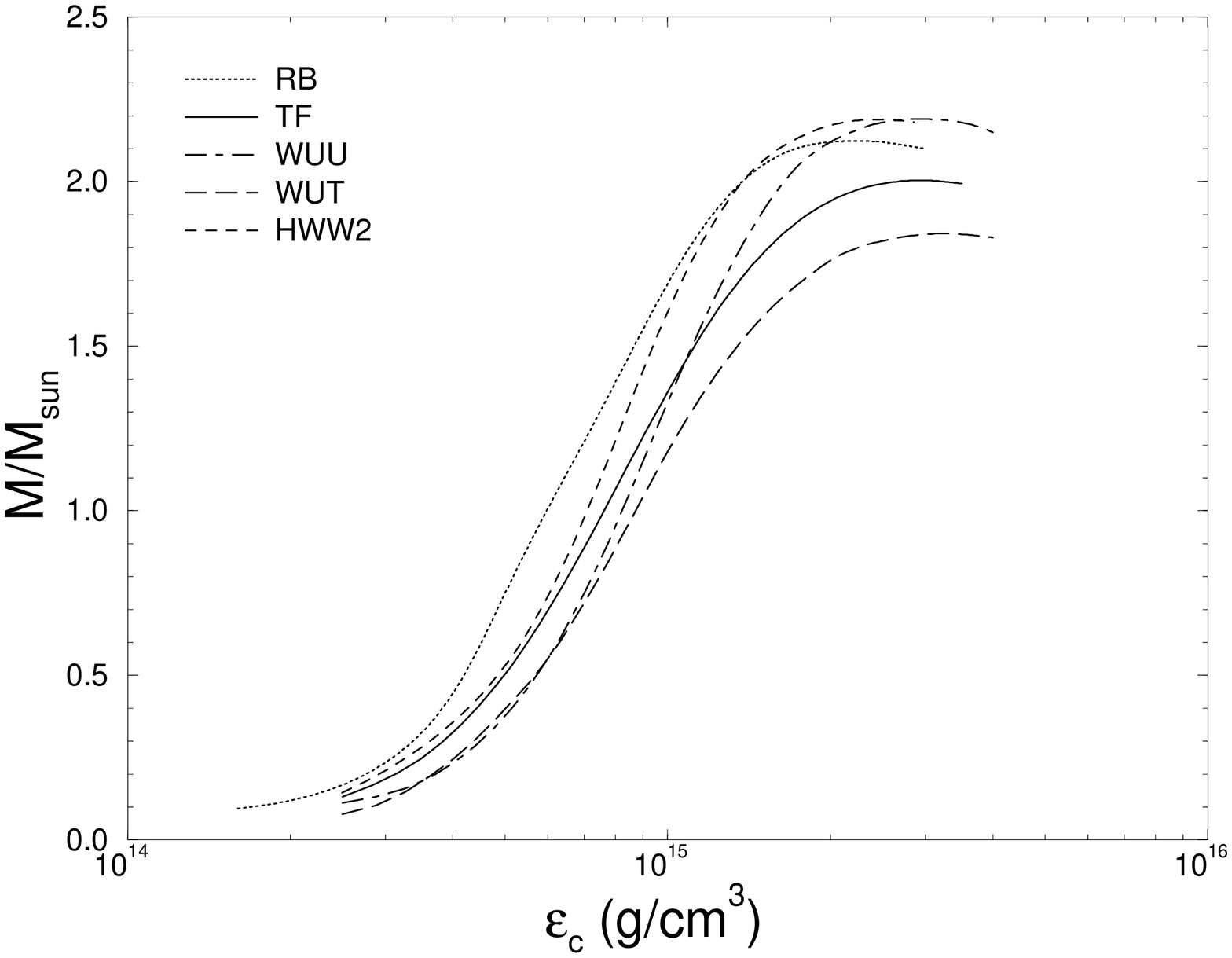,width=70mm,clip=,angle=-90,
          bbllx=95,bblly=35,bburx=565,bbury=680}
   \end{center}
   \end{figure}
   
   \begin{figure}[tbp] 
   \begin{center}  
   \leavevmode 
   FIGURE 11\\[0.5cm]
   \psfig{figure=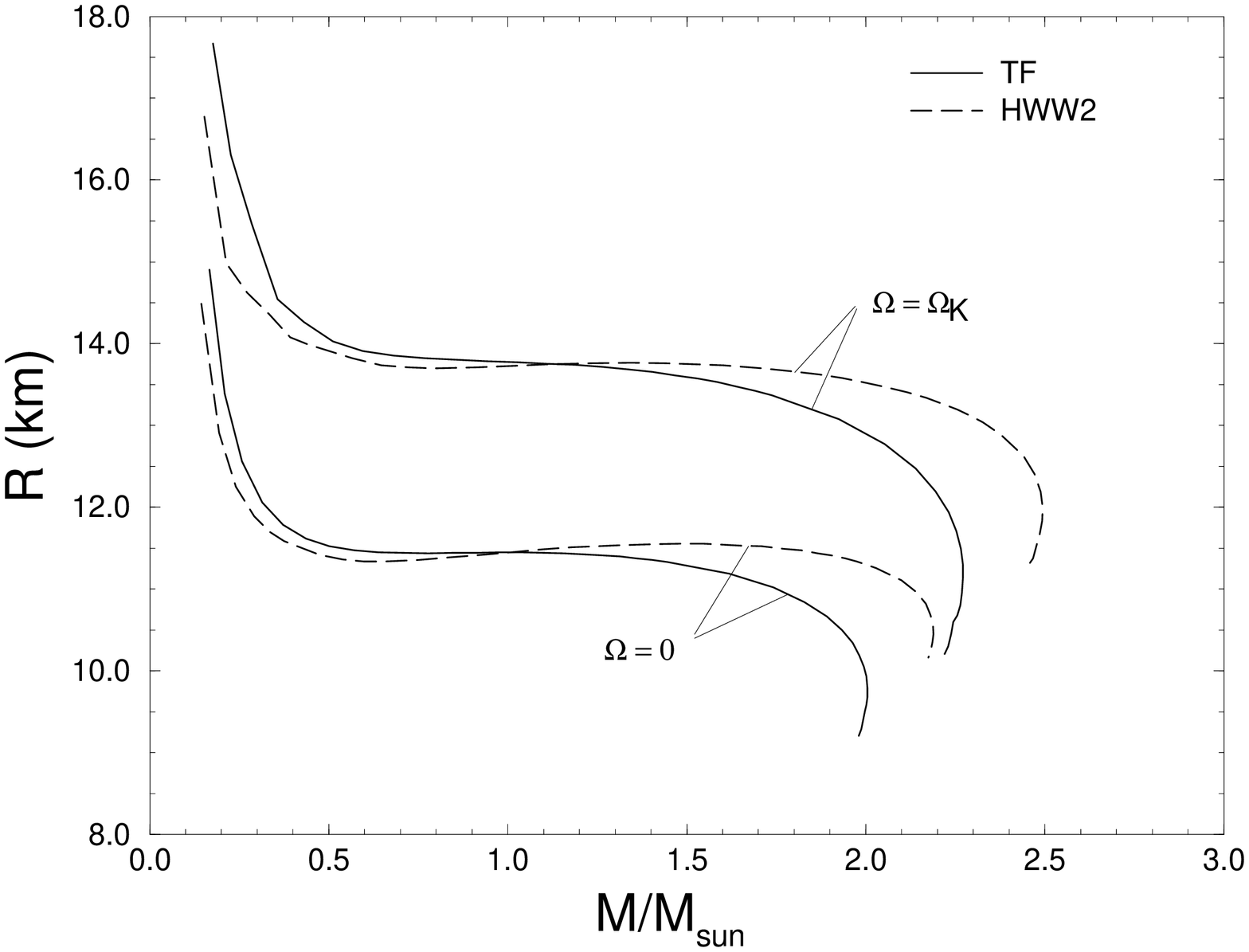,width=70mm,clip=,angle=-90,
           bbllx=95,bblly=35,bburx=565,bbury=680}
   \end{center}
   \end{figure}
   
   \begin{figure}[tbp] 
   \begin{center}  
   \leavevmode 
   FIGURE 12\\[0.5cm]
   \psfig{figure=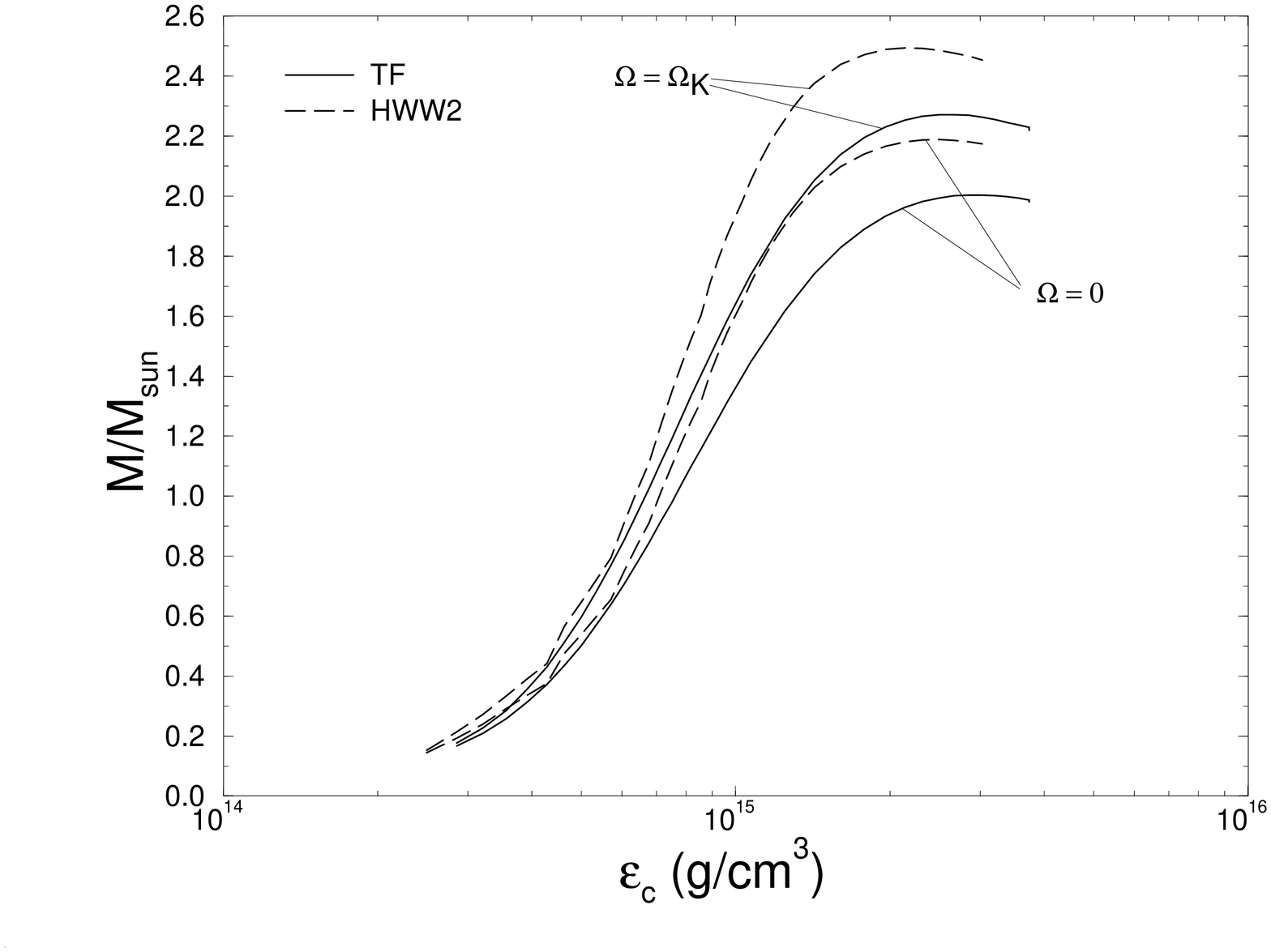,width=70mm,clip=,angle=-90,
           bbllx=95,bblly=35,bburx=565,bbury=680}
   \end{center}
   \end{figure}

   \begin{figure}[tbp] 
   \begin{center}  
   \leavevmode 
   FIGURE 13\\[0.5cm]
   \psfig{figure=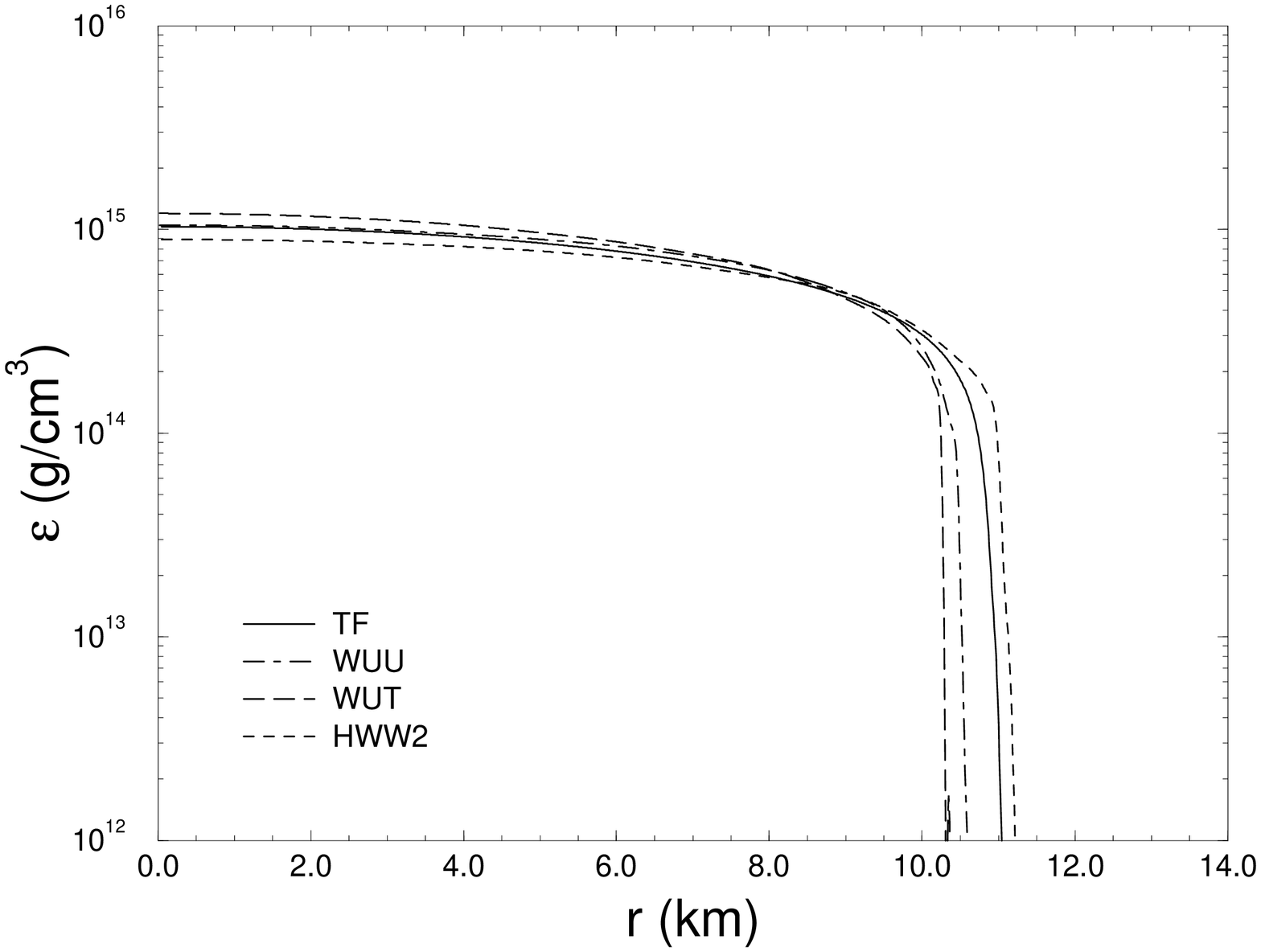,width=70mm,clip=,angle=-90,
          bbllx=95,bblly=35,bburx=560,bbury=680}
   \end{center}
   \end{figure}
   
   \begin{figure}[tbp] 
   \begin{center}  
   \leavevmode 
   FIGURE 14\\[0.5cm]
   \psfig{figure=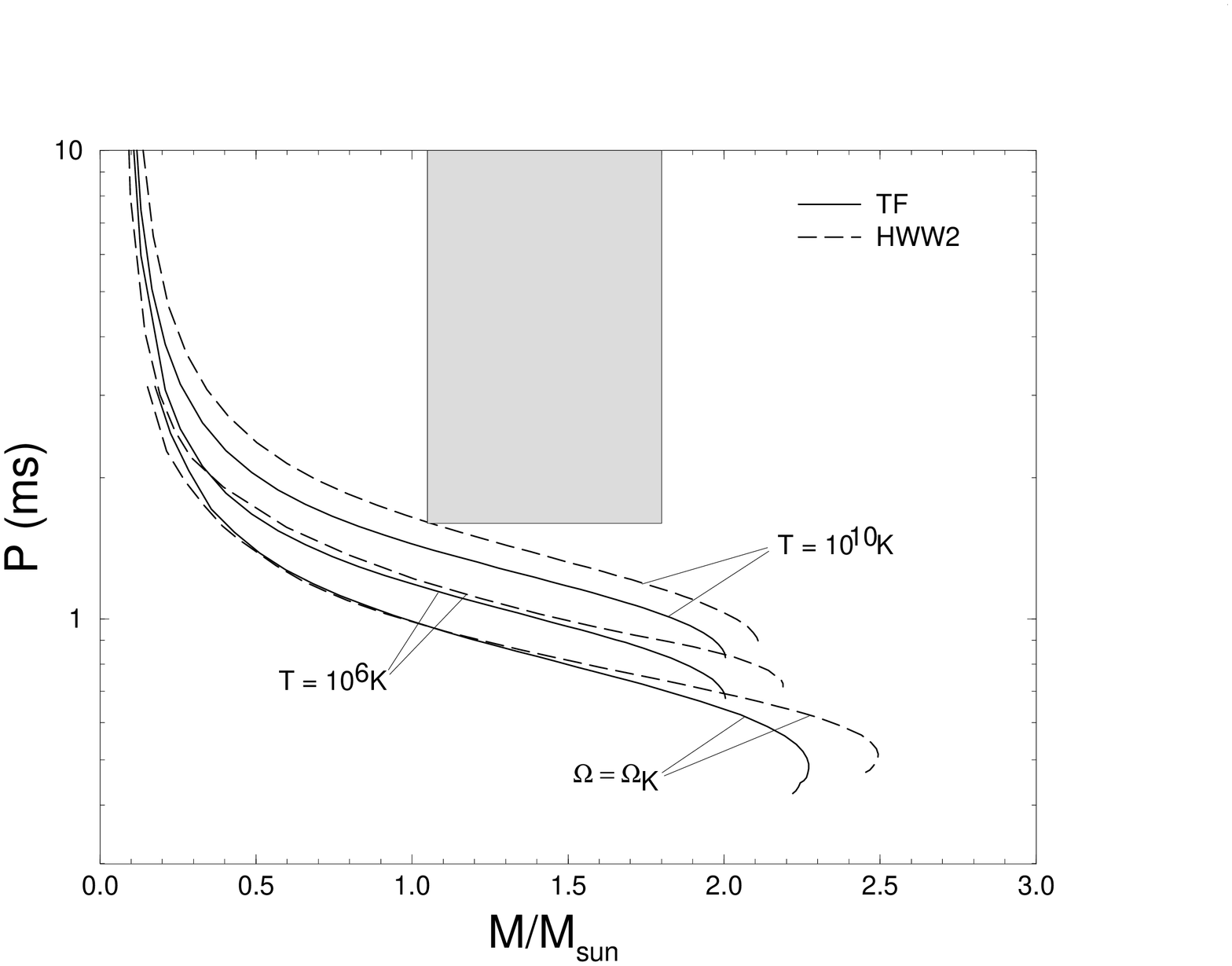,width=70mm,clip=,angle=-90,
          bbllx=95,bblly=35,bburx=565,bbury=680}
   \end{center}
   \end{figure}

   \begin{figure}[tbp] 
   \begin{center}  
   \leavevmode 
   FIGURE 15\\[0.5cm]
   \psfig{figure=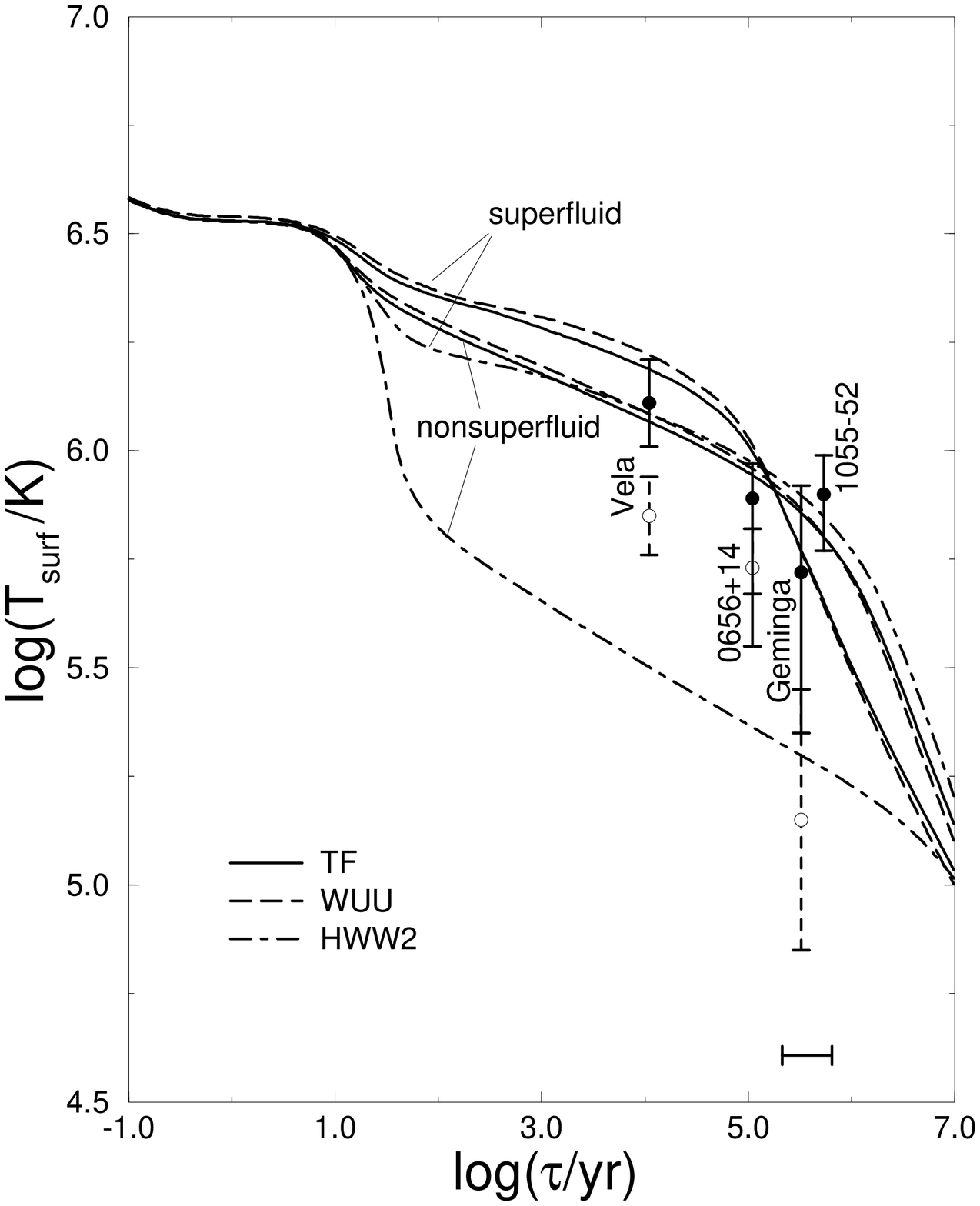,width=70mm,clip=,angle=0,
          bbllx=30,bblly=78,bburx=520,bbury=688}
   \end{center}
   \end{figure}